\documentclass[letterpaper,onecolumn,11pt,noarxiv]{quantumarticle}
\usepackage{amsmath,amsthm,amsfonts,amssymb,bbm,graphicx,mathtools,physics,thmtools,thm-restate}
\usepackage{changepage}
\usepackage{xprintlen}
\usepackage[utf8]{inputenc}
\usepackage[english]{babel}
\usepackage[T1]{fontenc}
\usepackage{mathrsfs}
\usepackage{comment}
\usepackage{algorithm}
\usepackage{algorithmicx}
\usepackage{algpseudocode}
\usepackage{xcolor}
\usepackage{hyperref}
\usepackage{complexity}
\usepackage{dsfont}
\usepackage{enumitem}

\hypersetup{
  colorlinks=true,
  hypertexnames=false,
  linktocpage,
  colorlinks=true, 
  urlcolor=magenta!90!black,    
  linkcolor=blue!60!black, 
  citecolor=black!60 
}
\usepackage[capitalize, compress]{cleveref}
\usepackage[numbers,sort&compress]{natbib}

\newclass{\stoqma}{StoqMA}
\newclass{\classP}{P}
\newclass{\bqp}{BQP}
\newclass{\qcam}{QCAM}
\newclass{\postbqp}{postBQP}
\newclass{\posta}{postA}
\newclass{\postiqp}{postIQP}
\newclass{\classa}{A}
\newclass{\bpp}{BPP}
\newclass{\fbpp}{FBPP}
\newclass{\pp}{PP}
\newclass{\cocp}{coC_=P}
\newclass{\ph}{PH}
\newclass{\np}{NP}
\newclass{\conp}{coNP}
\newclass{\gapp}{GapP}
\newclass{\approxclass}{Apx}
\newclass{\gapclass}{Gap}
\newclass{\sharpP}{\#P}
\newclass{\ma}{MA}
\newclass{\am}{AM}
\newclass{\qma}{QMA}

\newclass{\hog}{HOG}
\newclass{\quath}{QUATH}
\newclass{\bog}{BOG}
\newclass{\xeb}{XEB}
\newclass{\xhog}{XHOG}
\newclass{\xquath}{XQUATH}
\newclass{\maxcut}{MAXCUT}
\newclass{\sat}{SAT}
\newclass{\maxtwosat}{MAX2SAT}
\newclass{\twosat}{2SAT}
\newclass{\threesat}{3SAT}
\newclass{\sharpsat}{\#SAT}
\newclass{\se}{Sign Easing}
\newclass{\classx}{X}

\newtheorem{theorem}{Theorem}

\newtheorem{conjecture}[theorem]{Conjecture}
\newtheorem{definition}[theorem]{Definition}
\newtheorem{lemma}[theorem]{Lemma}

\newtheorem{corollary}[theorem]{Corollary}

\newcommand{\mc}{\mathcal}
\newcommand{\mb}{\mathbb}

\newcommand{\ix}{{\sf I}}
\newcommand{\sx}{{\sf S}}

\newcommand{\xx}{{\sf X}}

\newcommand{\tn}[1]{^{\otimes #1}}

\newcommand{\ee}{\mathrm{e}}

\newcommand{\bin}{\{0,1\}}

\newcommand{\proj}[1]{\ket{#1}\bra{#1}}

\newtheorem{fact}{Fact}

\newcommand{\Rank}[1]{\mathrm{Rk}\!\left(#1\right)}

\title{Random regular graph states are complex at almost any depth}
\author{Soumik Ghosh}
\affiliation{University of Chicago}
\author{Dominik Hangleiter}
\affiliation{Simons Institute for the Theory of Computing, University of California at Berkeley}
\author{Jonas Helsen}
\affiliation{QuSoft and CWI, Amsterdam, The Netherlands}
\date{\today}

\begin{document}

\begin{abstract}
  Graph states are fundamental objects in the theory of quantum information due to their simple classical description and rich entanglement structure. They are also intimately related to IQP circuits, which have applications in quantum pseudorandomness and quantum advantage. For us, they are a toy model to understand the relation between circuit connectivity, entanglement structure and computational complexity. In the worst case, a strict dichotomy in the computational universality of such graph states appears as a function of the degree~$d$ of a regular graph state [GDH+23]. In this paper, we study the average-case complexity of simulating random graph states of varying degree when measured in random product bases and give distinct evidence that a similar complexity-theoretic dichotomy exists in the average case. Specifically, we consider random $d$-regular graph states and prove three distinct results: First, we exhibit two families of IQP circuits of depth $d$ and show that they anticoncentrate for any $2 < d =  o(n^{1/2})$ when measured in a random $X$-$Y$-plane product basis. This implies anticoncentration for random constant-regular graph states. Second, in the regime $d = \Theta(n^c)$ with $c \in (0,1)$, we prove that random $d$-regular graph states contain polynomially large grid graphs as induced subgraphs with high probability. This implies that they are universal resource states for measurement-based computation. Third, in the regime of high degree ($d\sim n/2$), we show that random graph states are not sufficiently entangled to be trivially classically simulable, unlike Haar random states. Proving the three results requires different techniques---the analysis of a classical statistical-mechanics model using Krawtchouck polynomials, graph theoretic analysis using the switching method, and analysis of the ranks of submatrices of random adjacency matrices, respectively. 
\end{abstract}
\maketitle
\newpage
\setcounter{tocdepth}{2}
\tableofcontents
\section{Introduction}

Graph states play a fundamental role in the theory of quantum computation and communication \cite{hein2006entanglementgraphstatesapplications} as well as the complexity of physical systems \cite{raussendorf_computationally_2018}. 
They are arguably the simplest quantum states, with a classical description in terms of simple graphs, exhibiting rich quantum phenomena. 
From the perspective of multipartite quantum communication, they are interesting because local operations can transform the global graph topology and thus allow for flexible routing \cite{van_den_nest_graphical_2004,dahlberg_how_2020,hahn2019quantum}.
From the perspective of many-body physics, they are interesting since they relate to computationally distinct phases of matter~\cite{raussendorf_computationally_2018}.

In our work, we consider graph states from the perspective of understanding quantum properties that lead to computational speedups. 
Graph states are prime candidates to study a specific quantum phenomenon, namely, entanglement, in terms of how it relates to computational complexity. 
On the one hand, measuring graph states in adaptive single-qubit bases allows the execution of arbitrary quantum computations through measurement-based quantum computing~\cite{raussendorf_one-way_2001,raussendorf_measurement-based_2003}. 
In fact, these measurements can be restricted to the $X$-$Y$ plane of the Bloch sphere~\cite{mantri2017universality}. 
On the other hand, graph states exhibit a rich multipartite entanglement structure~\cite{hein_multiparty_2004} which is in fact required for quantum speedups~\cite{nest_classical_2007}. 
Understanding which properties of graph states makes them generically hard to simulate classically can therefore yield insights into the mechanisms underlying quantum speedups.

An insightful model to study these properties is the family of \emph{regular graph states}, i.e., graph states whose underlying graph on $n$ vertices is $d$-regular for some $0 \le d \le n$. 
These states form a family of graph states with a well-controlled connectivity structure, which relates to their entanglement and classical simulability.
The degree of the underlying graph also connects to the circuit depth required to prepare them, since they are prepared by applying controlled-phase gates to an initial $\ket{ +}^{\otimes n}$ state---all $d$-regular graph states can be prepared in depth at most $d+1$.
$d$-regular graph states are therefore amenable to preparation on near-term devices with long-range connectivity such as reconfigurable atom arrays \cite{bluvstein_logical_2024} and trapped ions \cite{decross_computational_2024}.
In fact, recent experiment demonstrated quantum advantage based on random universal circuits on regular graphs~\cite{decross_computational_2024}. 
In contrast to universal random circuits, graph states can be implemented via naturally fault-tolerant operations in certain stabilizer codes \cite{paletta_robust_2023,hangleiter_fault-tolerant_2025} and are therefore amenable to early fault-tolerant implementations \cite{bluvstein_logical_2024} as well as  tailored error mitigation \cite{martiel_low-overhead_2025}. 
Graph state preparations can also be efficiently verified  using high-quality single-qubit measurements~\cite{ringbauer_verifiable_2024}.
Thus, random regular graph states enable compelling experiments demonstrating noise-robust quantum advantage \cite{bergamaschi_quantum_2024}, 
efficient verification and benchmarking of generic structured circuits in long-range-connected architectures \cite{hangleiter_fault-tolerant_2025},
and first experimental preparations of highly structured states with applications beyond quantum advantage.\footnote{Examples of such applications are secret sharing \cite{markham_graph_2008} and multiparty computation \cite{kashefi_multiparty_2017}, see Ref.~\cite{hein2006entanglementgraphstatesapplications} for an early overview}

In the \emph{worst case} \cite{ghosh_complexity_2023} gave a tight connection between simulability and entanglement:
when measured in an arbitrary $X$-$Y$ plane product basis, $d$-regular graph states are hard to simulate and highly entangled if and only if $2 < d < n-3$~.
But both for demonstrating quantum advantage, and when aiming to understand the intrinsic relation between hardness of simulation, multipartite entanglement, and device connectivity, it is crucial to study \emph{generic} states from the family, i.e., random $d$-regular graph states \cite{hangleiter_computational_2023}.\footnote{Random instances from this family can be efficiently sampled~\cite{gao_uniform_2021}. }
This ensures that we are not drawing conclusions from isolated points in the family and makes random regular graphs a goldilocks model to study those fundamental and practical questions alike.\footnote{
Another, unrelated, setting in which random graph states arise naturally is measurement-based computation based on cluster states \cite{raussendorf_one-way_2001} with iid. erasure noise on all qubits \cite{browne_phase_2008}, or non-deterministic gates \cite{kieling_percolation_2007}. 
In these scenarios there are percolation thresholds between  easy-to-simulate and hard-to-simulate phases. 
}





In this paper, we therefore consider the \emph{average-case complexity}  of simulating uniformly random $d$-regular graph states when measured in an arbitrary $X$-$Y$ plane product basis. 
In doing so we study structured randomness in order to understand the relation between entanglement and complexity. 
From a fundamental perspective, this provides an alternative to fully random quantum states. 
From an experimental perspective, it provides an opportunity for demonstrating the computational capabilities of quantum processors for more structured circuits, as well as the potential for fault-tolerant implementations.
Importantly, the setting we consider covers both sampling problems and universality for measurement-based quantum computing.
We give evidence for the average-case complexity of these two problems in three distinct regimes of the regularity parameter, (1) the regime of $2 < d =  O(1)$, (2) the regime of $d \in \Theta(n^c)$ for any $1/2 < c <1$, and (3) the regime of $d \sim n/2$. 
To the best of our knowledge, we thus give the first ensemble of circuits which are classically intractable for \emph{any depth} (in particular low depth) above a constant threshold.
The best lower bound we had for such a threshold has been logarithmic depth~\cite{bremner_achieving_2017,napp_efficient_2022,hangleiter_fault-tolerant_2025,schuster_random_2024}. 

We first give evidence for average-case complexity in the regime of $2 < d = O(1)$. 
To this end, we first show that the output distributions of IQP circuits on random highly connected graphs constructed from $d$ random matchings measured in a random $X$-$Y$ plane basis have the anticoncentration property \cite{aaronson_computational_2013,hangleiter_anticoncentration_2018} at any depth $2 < d = o(n^{1/2})$. 
Combined with the fact that for each value of $d$ there are worst-case-hard instances in this family, this provides evidence for hardness of simulation on the same level of rigor as is known for other discrete families of IQP circuits \cite{bremner_achieving_2017,hangleiter_computational_2023}. 
This is the first family of quantum circuits that we are aware of for which there is evidence of simulation hardness at any sublinear depth above a constant threshold, and thus a result of independent interest.
We use this result to characterize the complexity of random regular graph states, noting that a random circuit from this matching model (strictly speaking a slight variation called the pairing model) at a fixed depth $d$ yields a uniformly random $d$-regular graph state with probability $O(2^{-d^2})$. This implies the anticoncentration property of constant-regularity graph states and thus gives evidence for their classical intractability, showing the first main result.

In our second result, we show that a random $d$-regular graph with $d = \Theta(n^c)$ for any $1/2 < c < 1$ has a grid graph of polynomial size as an induced subgraph (resulting from deleting some of the vertices). 
Since the grid graph is a universal resource for measurement-based quantum computation, this shows that an algorithm for sampling from the output distribution of those graph states in \emph{any} local basis would imply a collapse to the polynomial hierarchy up to the average-case \#P-hardness of a certain approximation problem (as well as being, in principle, universal resources for MBQC). To the best of our knowledge, this is the strongest known average case hardness result for any ensemble of graph states and the corresponding ensemble of IQP circuits.

Finally, in our third result, we show that uniformly random graph states have geometric entanglement bounded significantly away from the maximum. 
This makes them nontrivial and in particular not amenable to the trivial simulation algorithm of Ref.~\cite{gross_most_2009}, giving some evidence for their average-case complexity when measured in any local basis. We believe similar results will hold for random $cn-$regular graphs with $0<c<1/2$.

Altogether, our results for the average-case behavior turn out to be qualitatively similar to the worst-case behavior. 
However, they are rather more difficult to obtain, and substantial technical work using entirely different tools for each of the results is required, pointing to different properties of the three regimes: 
The first result requires the analysis of a statistical-mechanics mapping \cite{hangleiter_fault-tolerant_2025} which we show can be reduced to asymptotic properties of the Krawtchouk polynomials \cite{krawtchouk_sur_1929} which have previously appeared in a variety of contexts~\cite{feinsilver2005krawtchouk}. 
The second result requires graph-theoretic tools to show properties of subgraphs of random $d$-regular graphs. 
And the final result requires the analysis of the geometric entanglement entropy, which can be reduced to the study of extremal probability problems related to the rank of submatrices of uniformly random adjacency matrices.

Our results demonstrate that computationally universal or complex states can arise naturally from constrained randomness, and that this constraint can in fact give rise to more complexity than less structured or completely unstructured randomness. 
Indeed, comparing to the paradigmatic setting of quantum circuits composed of parallel Haar-random 2-qubit gates, two of our results are particularly striking. 
First, our result that random graph states which can be prepared in constant depth exhibit anticoncentration is provably not true for constant-depth Haar-random circuits \cite{Dalzell_2022, Deshpande_2022, schuster_random_2024}. These may still be average-case hard to simulate, but numerical evidence in low dimensions points against a low depth threshold for classical intractability \cite{napp_efficient_2022}. 
Second, our results that uniformly random graph states are not too entangled to be useful for measurement-based computation is also provably not true for Haar-uniformly random states \cite{gross_most_2009}.
Thus, our results suggests that while quantum states generated by Haar-random circuits are only complex in a limited depth regime, graph states generated by random CZ circuits are complex at almost any depth. 

\subsection{Guide for readers}

Although the paper is rather long, it is composed of three parts, each covering a different main result in a fairly self-contained manner.
Depending on which result they want to read about, the reader may jump over to the relevant section without needing to read the other parts of the paper. 
\begin{itemize}

\item \textbf{Anticoncentration results for IQP circuits and constant-degree regular graph states}: A summary of the results and proof techniques is given in \Cref{anticoncentration1}, \Cref{anticoncentration2}, \Cref{low}. The results are discussed in detail and proven in \Cref{sec:constant} with preliminaries in \cref{ssec:graphs prelim,subsec:models,ssec:krawtchouk prelims}. 
\item \textbf{Universality results for random regular graph states of intermediate degree}: A summary of the results and proof techniques is given in \Cref{universality1,intermediate}. The results are proven in  \Cref{sec:intermediate}.

\item \textbf{Absence of a geometric entanglement barrier for random graph states of high degree}: The main results and proof techniques are summarized in  \Cref{geometry2,high}. The results are proven in \Cref{sec: high degree} with preliminaries in \cref{ssec:markov prelims}.
\end{itemize}
Not to be missed in either case is the discussion in \cref{sec:discussion}.

\subsection{Results}
We present results on uniformly random $d$-regular graph states covering three different ranges for the regularity parameter $d$. We deal, in order, with $2 < d = O(1)$, $d = \Theta(n^c)$ for $c \in (0,1)$, and $d\sim n/2$. Along the way we will also prove several results for different random graph models, such as the pairing and matching models (described in \Cref{subsec:models}) and the uniformly random graph model. We believe these results to be of independent interest.

\subsubsection{Anticoncentration of a family of random pairing and matching IQP circuits}
\label{anticoncentration1}

Our first set of results deals with the average-case hardness of two families of IQP circuits of depth $d$ satisfying $d >2$ and $d = o(n^{1/2})$. 
These IQP circuits prepare (not always regular) graph states of degree at most $d$. 
Average-case hardness of constant-regularity graph states will follow from those results. 
To give evidence for average-case hardness of these depth-$d$ circuits we prove that their outcome distribution, when measured in a random local basis in the $X$-$Y$ plane \emph{anticoncentrates}. Moreover, they contain worst-case hard instances.

We will state our theorems in terms of the (normalized) second moment
\begin{equation*}
 m_2(G, \theta)  = 2^{n}\sum_{x\in \{0,1\}^n} p_{G, \theta}(x)^2 .
\end{equation*}
of the outcome distribution $p_{G,\theta}$ of the graph state on $G $ measured in $X$-$Y$-plane angles $\theta \in [0,2\pi)^{n}$. 
We say that the output distribution anticoncentrates if $m_2 \in O(1)$, since in that case a constant fraction of the output probabilities must be on the order of $1/2^n$.
Averaging over graphs $G$ and measurement angles $\theta$, the average second moment $\overline m_2 = \mb E_{G,\theta} [m_2(G, \theta)]$ gives evidence for the \sharpP-hardness of approximating the outcome probabilities of measuring states corresponding to random graphs $G$ and random angles $\theta$ up to constant relative error~\cite{hangleiter_computational_2023}. 
Intuitively, anticoncentration prohibits an average case simulator which uses trivial approximations by zero to most probabilities of most graph states in its simulation.
In this sense, it gives evidence for the average-case complexity of random graph states. 
Assuming additional well-founded complexity theoretic assumptions, in conjunction with the worst-case hardness results of \cite{ghosh_complexity_2023} this gives evidence for the hardness of sampling from such graph states to the same level of confidence as we have for other discrete families of circuits. The technical argument is reviewed in Ref.~\cite{hangleiter_computational_2023} and, roughly speaking, goes as follows: The circuit family we consider contains instances that are provably hard to strongly simulate up to relative error. Anticoncentration shows that most outcome probabilities have a similar order of magnitude and hence there is no detectable structure in the output distribution that would make a trivial simulation algorithm work. 
And finally, the ensembles we consider do not appear to have any exploitable structure that would help an algorithm designer to simulate a random instance compared to the worst-case instance, see also \cite{bremner_average-case_2016}.

Specifically, we consider the following two ensembles of depth-$d$ IQP circuits, which generate graph states of degree $d$. 
We call the first ensemble the \emph{random pairing model}. This model is obtained by choosing a uniformly random matching on $n\cdot d$ vertices and identifying $d$ vertices with a qubit. A $CZ$ gate is applied to every edge in the resulting multigraph. 
This model (which is well studied in graph theory~\cite{wormald1984generating}) is motivated by the fact that, conditioning on the graph being simple, it generates a uniformly random $d$-regular graph state. 
We associate (simple) graph states to the multigraphs chosen in these ways by deleting double edges and self-loops. 
In the second model, the \emph{random matching model}, a graph state is generated by applying $CZ$ gates on $d$ independent, uniformly random matchings. 
In this model, for any constant $d$, a regular graph state is generated with constant probability.
However, not all $d$-regular graph states admit a perfect matching \cite{janson_random_1995} (and some admit many, making this distribution different from the uniformly random regular graphs, even when conditioning on simplicity). 
Thus, the regular graph states generated by perfect matchings are exactly those graph states with $dn/2$ edges which can be prepared in optimal depth~$d$.

\begin{figure}
\centering
    \includegraphics{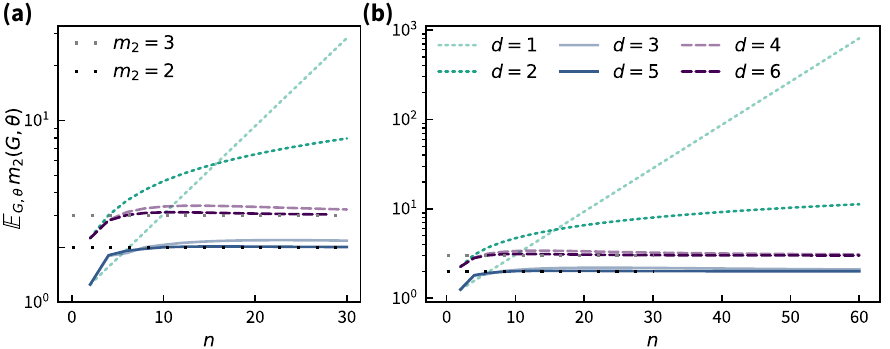}
    \caption{\label{fig:matching anticoncentration}
    Exact values of $\mb E_{G, \theta}[m_2(G, \theta)]$ (evaluated via \Cref{eq:anticoncentration random matching}) for graph states drawn from \textbf{(a)} the random pairing ensemble $\mc G_p(d,n)$ and \textbf{(b)} the random matching ensemble $\mc G_m(d,n)$ for various values of $d$ as a function of $n$. The opaque (transparent) loosely dotted lines denote the asymptotic values of $m_2 = 2(3)$ of the average second moment. Solid (dashed) lines represent odd (even) values of $d$. }
\end{figure}

We show that the output distribution of both models in a random $X$-$Y$ basis anticoncentrates for any $d $ satisfying $d > 2$ and $ d  = o(n^{1/2})$. 
\begin{theorem}[name=Anticoncentration of random pairing and  matching graph states,label=thm:anticoncentration]
Consider random graph states  $\ket G$ on $n$ vertices drawn from (a) the random pairing model on $nd$ vertices or (b) $d$ uniformly random matchings on $n$ vertices.
Then for $2 < d =o(n^{1/2})$
\begin{equation}
\mathbb{E}_{G, \theta}[m_2(G, \theta)]  = o(1) + \begin{cases}
    2 & \text{ if $d$ is odd }\\
    3 & \text{ if $d$ is even } .
\end{cases}
\end{equation}
\end{theorem}
We also show the exact value of $\mathbb{E}_{G, \theta}[m_2(G, \theta)] $ for different values of $n,d$ and both models in \Cref{fig:matching anticoncentration}. 
We observe not only that the $d$-dependence of our results is tight, but also that the convergence to the asymptotic value occurs rapidly.

The results of Ref.~\cite{ghosh_complexity_2023} imply that for each value of $d$ the random pairing model contains worst-case hard graphs. 
The random matching model trivially contains such graphs since for every odd $d$ it contains the hexagonal lattice, and for every even $d$ it contains the square lattice, both of which are universal for measurement-based quantum computation \cite{van_den_nest_universal_2006}. 
Thus, anticoncentration gives evidence for the average-case hardness of approximating the outcome probabilities in both cases. 
To the best of our knowledge, \Cref{thm:anticoncentration} is thus the first to give rigorous evidence for the hardness of simulating constant-depth circuits with random connectivity. 
In particular, it is the first result for constant-depth circuits which goes beyond resource states for measurement-based quantum computing (MBQC) \cite{terhal_adaptive_2004,gao_quantum_2017,bermejo-vega_architectures_2018,haferkamp_closing_2020}. 
At constant degree, we do not believe that random regular graphs  contain (as induced subgraphs) large 2D graphs (with high probability) which are usually required in MBQC constructions \cite{van2006universal}.

\subsubsection{Anticoncentration of constant-degree regular graph states}
\label{anticoncentration2}
Finally we can use our results on the pairing model to address the average-case complexity of random regular graph states of constant degree $d$.
Specifically \Cref{thm:anticoncentration} implies that random regular graphs with \emph{any constant degree} $d > 2$ anticoncentrate, giving evidence for their average-case hardness. This statement follows immediately from the following more general corollary
\begin{corollary}[name=Anticoncentration of random $d$-regular graph states,label=cor:anticoncentration]
\label{thm:random regular}
Consider the uniform measure $\mc G_r(n)$ on $d$-regular graphs on $n$ vertices, and uniformly random angles $\theta$. Then, for $2< d = o(n^{1/2})$, 
\begin{equation}
\mb E_{G \sim \mc G_r(d,n), \theta} [m_2(G,\theta)]  \leq (3+o(1)) 2^{d^2}.
\end{equation}
\end{corollary}
This corollary is a straightforward consequence of the fact that conditioning on simplicity in the pairing model gives rise to uniformly random regular graphs (see \Cref{subsec:models}).
\Cref{thm:random regular}, together with the worst case hardness result of \cite{ghosh_complexity_2023} gives evidence that approximating the outcome probabilities of regular graph states of constant degree measured random $X$-$Y$ plane bases is \sharpP-hard on average. We strongly believe that random regular graph states also anticoncentrate at super-constant depth, but leave showing that to future work. 

We note that some existing results about IQP circuits can be rephrased in terms of random graph state models yielding results similar to \cref{thm:random regular} for different models of random graphs.
In particular, Ref.~\cite{hangleiter_fault-tolerant_2025} shows that IQP circuits  composed of $g$ $CZ$  and $Z$ gates applied uniformly at random to $n$ qubits anticoncentrate if $g = \Omega(n \log n )$ but fail to anticoncentrate for any $g = O(n)$. 
This directly translates to the Erd{\H o}s-R\'enyi random graph model in which an edge is contained in the graph with probability $p = g/\binom n 2$, measured in the $\pm X$ basis. 
Ref.~\cite{hangleiter_fault-tolerant_2025} gives evidence that Erd{\H o}s-R\'enyi random graphs are hard to simulate for $ p \in \Omega(\log n /n)$ and suggests that they are not generically hard to simulate below that threshold. 
This is consistent with graph percolation, where famously Erd{\H o}s and R\'enyi \cite{erdos_evolution_1960} showed that if $p < 1/2n$, a random graphs largest connected component has size $O(\log n)$ which implies that it is efficiently simulatable using tensor-network techniques \cite{nest_classical_2007}. 
 
\subsubsection{Universality of regular graphs of intermediate degree}
\label{universality1}
Our second set of results is about the universality of random graph states, when the regularity of the graph scales quite strongly with $n$. In particular, when $d = \Theta(n^c)$ for $c\in (0,1)$ we argue that large grid graphs can be found as induced subgraphs in random regular graphs with high probability. This implies that the associated graph states can be turned into universal resource states using only computational basis measurements. We have: 
\begin{theorem}[name=Induced grid graphs,label=thm:grid_subgraph]
Let $G$ be a random $d$--regular graph on $n$ vertices, with $d= n^c$ where $0.5 < c < 1$. Then, with probability $1 - o(1)$, it contains a square grid graph on $v$ vertices, for any $v= o(n^k)$, with $k = \mathsf{min}\left\{\frac{1-c}{2}, \frac{c}{3} \right\}$, as an induced subgraph.
\end{theorem}
This theorem is proven using the switching technique due to McKay and Wormald \cite{mckay1990uniform}. This is a standard technique in random regular graph theory, but is usually applied in situations where the subgraph to be found is of constant size (whereas for us it must grow reasonably fast with $n$). Because of this, and because these techniques are not widespread in the quantum computing literature, we give explicit switching calculations in \Cref{subsection: switching}.

A limitation of this result is the fact that $c>0.5$. This seems to be an unavoidable fact, as the expected number of grid graphs goes to zero whenever $c< 0.5$. This threshold behavior is observed even for grid graphs of constant size \cite{kim_small_2007}. We can get around this limitation by considering \emph{sparsified} grid graphs, constructing by replacing every edge in an $L\times L$ grid graph by $L$ edges and $L-1$ vertices in a line. The resulting graphs are very line-like asymptotically yet are still universal (with a polynomial space overhead): we can recover the $L\times L$ grid graph as a \emph{vertex minor} by measuring all the added vertices in the $Y$ basis. For these graphs we can prove the stronger statement: 
\begin{corollary}[name=Sparsified induced grid graphs,label=thm:grid_subgraph_sparsified]
Let $G$ be a random $d$--regular graph on $n$ vertices, with $d= n^c$ where $0.5 < c < 1$. Then, with probability $1 - o(1)$, it contains a sparsified square grid graph on $v$ vertices, for any $v= o(n^k)$, with $k = \mathsf{min}\left\{\frac{1-c}{2}, \frac{c}{3} \right\}$, as an induced subgraph.
\end{corollary}

\Cref{cor:expected_grid_sparse} implies that random $n^c$-regular graphs of intermediate degree $c \in (0,1)$ are universal resources for measurement-based quantum computing when measured in an arbitrary product basis. 
Note, however, that \emph{finding} the induced sparsified grid graph may be (and probably is) a computationally difficult problem. 
Therefore, the standard reduction from sampling to computing probabilities of post-selected BQP computations may not be possible in polynomial time. 
However, since identifying a grid graph can be done with access to an NP oracle, the reduction is possible in the polynomial hierarchy. 
Thus, while an efficient exact sampling algorithm in an arbitrary basis from a grid graph collapses the polynomial hierarchy to the third level \cite{hangleiter_computational_2023}, it still collapses it to the fourth level for graphs containing induced sparsified grid graphs.

Finally it must be noted that the proofs of the above results explicitly break down in the regime where $d = cn$ with $c\in (0,1)$. More strongly, it is known that in this regime one can not find large (much larger than $\log(n)$ sized) non-trivial \emph{induced subgraphs} with more than negligible probability \cite{rucinski1987induced}. This leads one to suspect that graph states of linear degree are not universal. We suspect that it is in fact the case that they are universal on average, but we have no proof of this. We do provide some evidence in this direction, which we discuss in the next section.

\subsubsection{Absence of a geometric entanglement barrier for uniformly random graph states} 
\label{geometry2}

Entanglement is usually considered to be a necessary condition for universal measurement based quantum computation. However it was proven in \cite{gross_most_2009} that a state can be `too entangled' to be used as a resource state (and also that---under the Haar measure---most states are too entangled in this way). This was done by arguing that if a resource state has high geometric entanglement, defined as
\begin{equation}
E_g(\ket{\psi}) =  -\log\left(\max_{\alpha \in \mathrm{PROD}_n}|\braket{\alpha}{\psi}|^2\right).
\end{equation}
where $\mathrm{PROD}_n$ is the set of product states, then any NP problem solved by an MBQC with $\ket{\psi}$ as a resource state can be solved by a classical computer (with access to randomness) in time $O\big(\mathrm{poly}(n) 2^{n- E_g(\psi)}\big)$. In light of the failure of the arguments above in the $d=\Theta(n)$ regime one can wonder whether this entanglement barrier shows up for random graph states (of degree $cn$ with $c\in(0,1)$). We give evidence that this is not the case by proving an upper bound on the geometric entanglement of uniformly random graph states.
\begin{theorem}[name=Geometric entanglement upper bound for graph states,label=thm:geom_ent_upper]
Choose a graph state $\ket G$ uniformly at random. There exists constants $c,C$ such that
\begin{equation}
\mathbb{P}\bigg[E_g(\ket{G}) \geq n-cn^{1/4}/\log(n) \bigg]\leq C.
\end{equation}
\end{theorem}
The proof of this theorem establishes a connection between the geometric entanglement of random graph states and the behavior of the ranks of the principal submatrices of random adjacency matrices, which can subsequently be analyzed using ideas from extremal probability and Markov chain Monte Carlo \cite{rosenthal_efficient_2023}.
This upper bound is strong enough to exclude the simulation algorithm given in \cite{gross_most_2009}, which now runs in time $\Omega(2^{n^{1/4}})$. We note that this upper bound does not directly translate to a similar bound on $d$-regular graphs with $d = cn$. However these distributions are very similar. For instance, the sandwich theorem~\cite{gao2020sandwiching} tells us that uniformly $cn$-regular graphs can be closely related to Erd{\H o}s-R\'enyi random graphs with constant acceptance probability $p$ (our result can be interpreted as evaluating the $p=1/2$ point. Furthermore, uniformly random graphs are with high probability quite regular, with most vertices having degree $n/2 \pm O(\sqrt{n})$. We leave making these connections fully rigorous for future work. \\

We also provide strong evidence that this upper bound is almost tight, by providing a lower bound on the geometric entanglement of random stabilizer states (which are equivalent to graph states up to local Clifford operations).
\begin{theorem}[name=Geometric entanglement lower bound for stabilizer states,label=thm:geom_ent_lower_stab]
Choose a stabilizer state $\ket S$  uniformly at random. There exists a constant $c$ such that
\begin{equation}
\mathbb{P}\bigg[E_g(\ket{S}) \leq n - c\sqrt{n}\log(n) \bigg]\leq O(2^{-\sqrt{n}}).
\end{equation}
\end{theorem}
This theorem is a straightforward application of the representation theory of the Clifford group. Because a constant fraction of stabilizer states are graph states (up to linear phases, which leave the geometric entanglement unchanged) this leads to a similar statement for graph states.
\begin{corollary}[name=Geometric entanglement lower bound for graph states,label=cor:geom_ent_lower_graph]
Choose a graph state $\ket G$ uniformly at random. There exists a constant $c$ such that
\begin{equation}
\mathbb{P}\bigg[E_g(\ket{G}) \leq n - c\sqrt{n}\log(n) \bigg]\leq O(2^{-\sqrt{n}}).
\end{equation}
\end{corollary}
We believe (ignoring the $\log(n)$ factors) that the $O(\sqrt{n})$ deviation in \Cref{cor:geom_ent_lower_graph} is in fact accurate, and the slightly weaker $\Omega (n^{1/4})$ scaling in \Cref{thm:geom_ent_upper} is a consequence of the proof technique.
We leave closing the gap between the upper and lower bounds for future work.

\subsection{Proof ideas }

We presented results in three different regimes of regularity. The proof techniques in these three regimes are all quite different, drawing on results in combinatorics, random graph theory and random matrix theory. Here we outline, organized by regime, the techniques used in this paper. 

\subsubsection{Constant-degree graph states and random pairing and matching IQP circuits}
\label{low}
The key result on graph states of constant degree is \Cref{thm:anticoncentration}, which concerns random IQP circuits generated from uniformly random matchings. 
This result implies anticoncentration of the $X$-$Y$ plane output distribution of  constant-degree regular graphstates.  
To show this theorem, we make use of a combination of techniques. 
First, we reduce the problem  to a purely graph-combinatorial problem using a statistical mechanics interpretation of the expected second moment of the output distribution described in \cite{hangleiter_fault-tolerant_2025}.
Importantly, this model is distinct from similar models for quantum circuits with Haar-random single-qubit gates \cite{zhou_emergent_2019,hunter-jones_unitary_2019,dalzell_random_2022} and significantly more involved to analyze. 
The resulting combinatorial problem, which amounts to counting the number of matchings which have an even number of edges crossing between two arbitrary subsets of vertices, can be further interpreted as a sum over so called Krawtchouck polynomials \cite{krawtchouk_sur_1929}. 
These have seen use in coding theory \cite{feinsilver2005krawtchouk}, and good bounds are available \cite{derksen_pseudorandomness_2024}, which allow us (with a substantial amount of combinatorial elbow grease) to provide bounds on the expected second moment.

\subsubsection{Intermediate degree graph states}
\label{intermediate}

The main results on intermediate degree graphs (\Cref{thm:grid_subgraph,thm:grid_subgraph_sparsified}) are essentially about proving bounds on the appearance of induced subgraphs of random $d$-regular graphs. Proving such bounds is a well studied problem in graph theory, both in the case of $d$-regular graphs or Erd{\H o}s-R\'enyi graphs and both in the induced and standard subgraph cases (see e.g., \cite{kim_small_2007} or the book by Bollob\'as \cite{bollobas1998random}). 
However in the literature usually only the case of constant sized subgraphs is treated explicitly, whereas we require bounds for the appearance of induced subgraphs that grow quite fast with $n$. Thus, we prove \Cref{thm:grid_subgraph} by a careful application of existing combinatorial methods, in particular the method of switchings, introduced by McKay~\cite{mckay1990uniform}, which is particularly effective at analysing expectation values of random variables induced by random $d$-regular graphs, and the second moment method, to convert expectation value estimates to statements that hold with high probability. The difficulty here lies almost entirely in the care required to get nontrivial estimates for induced subgraphs growing in size with $n$. 

\subsubsection{High degree graph states}
\label{high}

The main results on graph states of high degree are \Cref{thm:geom_ent_upper,thm:geom_ent_lower_stab}, providing lower and upper bounds on the geometric entanglement. 
\Cref{thm:geom_ent_lower_stab} follows a relatively standard path, approximating the continuous optimization in the geometric entanglement by a discrete one through an epsilon net, followed by a union bound and a tail bound on the overlaps with fixed product states. There is some subtlety in that we require a rather small epsilon-net for the union bound to be non-trivial. To that end we extend a nice trick from random matrix theory to the geometric entanglement in \Cref{lem:eps_net_bound}, proving that we can obtain a multiplicative approximation to the maximum in the geometric entanglement using a relatively small epsilon net. The subsequent tail bound is then provided by the moments of random stabilizer states, developed in \cite{gross_schurweyl_2021}. Some care must be taken to choose the right moment here, as the moments of random stabilizer states grow too fast for a straightforward exponential generating function approach to work~\cite{helsen2023thrifty}.

Proving \Cref{thm:geom_ent_upper} is substantially more complicated. We restrict the optimization in the geometric entanglement to a special subset of product states for which we can characterize the overlap purely in terms of the corank of principal submatrices of the adjacency matrix of the graph state. This reduces bounding the geometric entanglement to an extremal probability problem on random symmetric binary matrices. Inspired by similar arguments in the theory of Gaussian processes~\cite{talagrand2014upper}, we then bound the correlation between the coranks of different principal submatrices of a single random adjacency matrix. We prove that if the overlap between the two matrices is small then their coranks are approximately independent. This is then enough to prove a lower bound on the maximal corank, via the second moment method (in particular we use the Bonferroni inequalities). Proving this approximate independence is done by reducing the problem to a Markov chain on the integers $\mathbb{N}$, for which we then prove precise (non-spectral) upper bounds on the convergence to the stationary state. 

\subsection{Context and prior work}
\label{sec:context}

Our work builds on first steps made in Ref.~\cite{ghosh_complexity_2023} that classify the simulation complexity and entanglement properties in the \emph{worst case} over the choice of $d$-regular graphs. 
There, it was shown that as $d $ is increased, the simulation complexity and entanglement properties undergo two phase transitions, providing a tight link between complexity and entanglement: For $d \le 2$ and $d \ge n-3$ the entanglement is low and simulating single-qubit measurements in any basis is classically easy for all $d$-regular graph, while for any other value of $d$ there is a $d$-regular graph for which simulations are classically intractable and the multipartite entanglement is high. 
But are these hard instances isolated in their respective regularity class or are \emph{most} instances of a class hard? In many cases, average-case complexity significantly differs from worst-case complexity, most famously so for \np-problems, where often most instances are in fact efficiently solvable.

Thus, our results on \emph{average-case hardness} of random $d$-regular graph states significantly strengthen the connection between the connectivity of a graph and the entanglement properties of the corresponding state and provide complexity-theoretic evidence that classical intractability is a generic feature of multipartite entangled states. Generally speaking, though, there are only few techniques to address average-case complexity such as random self-reducibility---these techniques primarily involve reducing the average-case problem to proving average-case hardness of computing the permanent of a matrix, for e.g. see \cite{Valiant1979, aaronson2010computationalcomplexitylinearoptics, Bouland_2018}---which work for Haar random gate sets since they are continuously parameterized but are not applicable to the discrete randomness of random graphs. 
This is why the average-case complexity of random graph states is a qualitatively different question from both their worst-case complexity, and average-case complexity of continuous ensembles. 

From a different perspective, our work can be viewed as a first attempt to answer the question whether random regular graph states are resources for measurement-based computing (MBQC) (and in the regime of $d =\Theta(n^c) $ we answer in the affirmative). 
Assessing which graphs form resource states for MBQC has a rich history~\cite{briegel_persistent_2001,raussendorf_measurement-based_2003,hein2006entanglementgraphstatesapplications,van_den_nest_universal_2006,Briegel_2009} and has motivated the idea of  computational phases of matter \cite{DohertyBartlett2009, ChungBartlettDoherty2009, Miyake2010, DarmawanBrennenBartlett2012, ElseSchwarzBartlettDoherty2012, ElseBartlettDoherty2012, RaussendorfYangAdhikary2023}. 
This question has also arisen when studying the impact of noise on computational power. 
Concretely, Browne \emph{et al.}~\cite{Browne_2008} studied the impact of erasure noise on MBQC on grid graphs in terms of percolation phenomena. Here, at each lattice site, there is a finite probability that the qubit on that lattice site is erased, resulting in a random cluster state as originally introduced by Briegel and Raussendorf \cite{briegel_persistent_2001}. 
Below a certain value of that probability---known as the percolation threshold---noisy grid graphs can then be classically simulated, whereas above that probability, noisy grid graphs remain universal resource states. 
Equivalently, we can view this setting as the random graph state ensemble given by random subgraphs of the grid graph \cite{briegel_persistent_2001}, which is a rather different ensemble than the ones studied in this paper.


\subsection{Discussion  and outlook}
\label{sec:discussion}

Our results are the first to explore the average-case complexity of random graph states with bounded degree. 
They give evidence towards the average-case complexity of simulating graph states of any degree $2< d \lesssim n/2$ 
using different types of results relating to the hardness and simulability of random graph states. 
In particular, our results interpolate between these extremely degree regimes.  
This interpolation runs via the degree of regular graphs and thus extends the results of \cite{ghosh_complexity_2023} to the average case.

However, they can by no means be said to be the final word on the average-case complexity of regular graph states, or low-depth quantum circuits. Thus, they  raise a number of interesting questions. 
In particular, they have interesting consequences when related to a variety of different themes in the study of simulating sampling from quantum circuits. 
In our discussion, we will discuss each regularity regime in turn and formulate a number of open questions and conjectures.

\subsubsection{Constant degree graph states and random matching IQP circuits}

Our \Cref{thm:anticoncentration} gives evidence for the average-case hardness of simulating family of random quantum circuits with random connectivities of any depth larger than $2$ and scaling slower than the system size $n$.
It is the first circuit family for which anticoncentration has been shown extending from any constant depth to sub-linear depth.  

Similar constructions of universal random circuits based on regular graphs with perfect matchings have previously been studied numerically and experimentally in \cite{decross_computational_2024}. 
Since commuting IQP circuits have often been precursors to results for random circuits (e.g. in terms of complexity \cite{bremner_average-case_2016,boixo_characterizing_2018} and noisy simulation \cite{bremner_achieving_2017,gao_efficient_2018,aharonov_polynomial-time_2022}), our result may thus also help in the rigorous study of constant-depth random circuits in arbitrary geometries. 
In particular, it is worth considering our results in relation to the results of \cite{napp_efficient_2022}, who give evidence that average-case hardness fails for constant-depth Haar-random circuits, as well as to the results of \cite{decross_computational_2024} who give evidence that the simulation complexity of random circuits on different connectivities remains bounded at very low depths. To see why our results are different, we observe that both of those works consider random circuits in which arbitrary single-qubit rotation gates can be applied throughout the circuit. 
It may indeed be that in this case, average-case hardness requires a strictly super-constant circuit depth. 
In fact, at least depth $\Omega(\log \log n)$ is required for anticoncentration in models that are invariant under Haar-random single qubit gates \cite{schuster_random_2024}. 
Our results circumvent this lower bound by restricting the local bases we measure in to the $X$-$Y$ plane.
Hence, the fact that we only measure in the $X$-$Y$ plane is critical to the anticoncentration in constant depth we show.
It remains an interesting question, however, how the simulation complexity of such random circuits depends on the circuit depth. 
As a first step towards this, it would be interesting to consider the random matching IQP circuits with measurements in a Haar-random single qubit basis. 
By the result of \cite{schuster_random_2024}, anticoncentration will fail for constant depth, and thus this model would be helpful to understand the mechanism governing the arising lower bound.

It is also interesting to consider the relation of our results to recent results for simulating noisy IQP circuits \cite{rajakumar_polynomial-time_2024}. There, it is shown that noisy IQP circuits at any depth larger than a noise-dependent constant can be efficiently classically simulated, and one might think that our results might yield some leeway to circumvent these results. 
However, note that a previous noisy simulation algorithm due to Bremner, Montanaro and Shepherd \cite{bremner_achieving_2017} will work for any IQP circuits which anticoncentrates, and therefore also applies to the circuit families we consider here. 
Noisy constant-depth IQP circuits might also be interesting to consider in the context of the recently discovered transition in the cross-entropy benchmark (XEB) versus the fidelity \cite{dalzell_random_2024,deshpande_tight_2022,gao_limitations_2024,ware_sharp_2023}. 
While IQP circuits giving rise to Erd{\H o}s-R\'enyi random  graph states with some edge probability $p$ exhibit the transition in the XEB \cite{hangleiter_fault-tolerant_2025},  this transition may be understood as occurring because in such graphs the probability that no gate is applied to a particular qubit vanishes only exponentially in $d$ rather than $nd $.
In contrast, for IQP circuits giving rise to $d$-regular graph states entangling gates are guaranteed to be applied to every qubit, and thus the transition may disappear.
This would allow them to be reliably and  (sample-)efficiently benchmarked using XEB.

\subsubsection{Intermediate degree graph states and universality for measurement-based computation}

Our results focus on the question of whether random graph states can be classically simulated, which is a priori a question which is distinct from the question whether random graph states are universal resources for measurement-based quantum computation. 
While some of our results (in the intermediate regime) explicitly make use of universal resources, it is interesting to ask in what sense or to what extent random graph states in different regimes can function as universal resource states. 
To answer this question, it must first be made clear that the notion of a ``universal resource'' can have very different meanings. 
A natural notion of universality is one in which we just ask that for any number of qubits $n$, the family of states contains a state on $m \ge n$ qubits such that using measurements and feed-forward an arbitrary unitary can be implemented on $n$ of the $m$ qubits up to some precision threshold~\cite{nest_universal_2006}. 
Note that this notion is not concerned with efficiency, and hence the graphs containing a 2D grid graph in the intermediate regularity regime are certainly examples of universal resources in this sense. 
The universal resource is even efficient in the sense that $m = \poly(n)$. 
However, note that these graphs cannot be efficiently found in general, and finding them may require the solution of an NP problem (showing this is an interesting open question). 
In order to computationally exploit a universal resource, it is a prerequisite that a sufficiently large 2D subgraph can also be efficienlty found, since the only known efficient constructions make use of those graphs.
This raises the question whether in the intermediate degree regime certain random ensembles contain grid graphs as subgraphs that can also be efficiently found.

In the regimes of very high and very low degree, whether or not random regular graph states are universal resources remains an interesting open questions. 
In particular, in those regimes, we do not expect there to be large grid-graphs or hexagonal graphs as induced subgraphs of many regular graphs. 
But all known ways of compiling a quantum computation in a measurement-based way make explicit use of the presence of such a subgraph. 
We believe the region in which explicit large grid graphs can be found, can be expanded (e.g. into the $d = cn$ regime) but only by going beyond induced subgraphs. Ideally, one would like to characterize which regular graphs have large grid graphs as \emph{vertex minors}, which is the graph-theoretic notion capturing the action of local Clifford gates and Pauli measurements. The vertex minor problem is NP-hard in the worst case \cite{dahlberg2022complexity}, but little is known about its average-case behavior. 
Even to understand whether grid graphs on an exponentially small, or even constant subset of qubits can be found remains an interesting open question.
There might also be resources for measurement-based quantum computation which are not equivalent to hexagonal or grid graphs. 
We think this is a fruitful avenue of future research. 

\subsubsection{High degree graph states}

In the regime of high degree we also lack a full characterization of the complexity of graph states, in particular the question of classical simulability is not fully settled. While we have excluded the simulation algorithm given in \cite{gross_most_2009}, we can not exclude the existence of more tailor-made classical algorithms for MBQC with random linear-degree graph states. In \Cref{subsec:structured} we described a simulator where all Pauli measurements are explicitly calculated and only non-Pauli measurement are simulated. We believe that even this more sophisticated algorithm fails, but can not prove it as of yet. We leave this as a conjecture. It would also be interesting to prove variants of our results for random $cn$-regular graph states with $0c<1/2$, which we believe can be done through the sandwich theorems from random graph theory \cite{gao2020sandwiching}.

\subsubsection{Towards analyzing noisy, architecture-constrained graph states}

Aside from the different regimes of the regularity parameter just discussed, an interesting open question raised by our work is the complexity of graph states that can be naturally realized in different quantum computing architectures. 
The arbitrary connectivity required to generate the random regular graphs may be difficult to realize in practice. 
For instance, while in principle reconfigurable atom or ion arrays \cite{bluvstein_logical_2024,decross_computational_2024} arbitrary connectivity can be realized, there may be more natural random ensembles of graph states, for instance, subgraphs of a high-dimensional hypercube~\cite{hangleiter_fault-tolerant_2025}. 
In other architectures, such as superconducting qubits, even more constrained low-dimensional lattice geometries are imposed~\cite{arute_quantum_2019}. 

A further interesting question in the context of more realistic circuit ensembles is to what extent noise affects the output distribution. 
In the literature on sampling from the output distributions of random quantum circuits, a prominent quantity is the cross-entropy benchmark (XEB) score, which generalizes anticoncentration to noisy circuits and serves as a measure of quality of the sampled distribution \cite{boixo_characterizing_2018,arute_quantum_2019,kliesch_theory_2021}.
An important result in this context has been that for local noise rates on the order of $ \lesssim 1/n$ and where the ideal output distribution anticoncentrates, the XEB can be used as a proxy of the many-body fidelity of the pre-measurement state \cite{dalzell_random_2024,morvan_phase_2024,ware_sharp_2023}, and moreover, that this is the regime in which there are no exploits that classical algorithms might use to achieve a high XEB score significantly more efficiently than brute-force simulations. In contrast, in the regime where the noise rate is larger than $1/n$, such exploits exist \cite{gao_limitations_2024}. 
In fact, as the depth is increased 
Similar statements can also be made for IQP circuits \cite{hangleiter_fault-tolerant_2025,bluvstein_logical_2024}, and it is an interesting open question to analyze the XEB for noisy random graph state ensembles. 

\section*{Acknowledgements}

JH would like to thank Ingo Roth and Jop Briet for useful suggestions on the proof of \Cref{thm:geom_ent_upper}, and Michael Walter and Marcel Hinsche for helpful discussions on the Clifford group. 
DH thanks Michael Gullans for many inspiring discussions on stat-mech mappings and IQP circuits. 
We also thank Daniel Grier, Jackson Morris, and Debbie Leung for discussions on vertex minors and simulation algorithms. 
We gratefully acknowledge the hospitality of the Simons Institute for the Theory of Computing in the spring of 2024, supported by DOE QSA grant \#FP00010905, where part of this work was conducted.
DH acknowledges funding from the Simons Institute for the Theory of Computing, supported by DOE QSA. 
JH acknowledges funding from the Dutch Research Council (NWO) through a Veni grant (grant No.VI.Veni.222.331) and the Quantum Software Consortium (NWO Zwaartekracht Grant No.024.003.037).

\section{Preliminaries}
\label{prelim}
In this section we recall some assorted facts about graphs, graph states and stabilizer states that we will need throughout the rest of the paper. This is by no means meant to be an exhaustive introduction. For graph theory (with a focus on random graphs) we recommend the classic textbook \cite{bollobas1998random}, and good introductions to graph states and stabilizer states can be found in \cite{nielsen_quantum_2010,hein_multiparty_2004}. We will also require properties of Krawtchouk polynomials as well as bounding techniques for the convergence of Markov chains, which we also recall in this section.

\subsection{Graphs, graph states and stabilizer states}
\label{ssec:graphs prelim}
We begin by reviewing some standard graph theoretic notions. 
A graph $G$ is a set of vertices $V$ (usually the set $[n]$, and a set of vertices $E\subset V\times V$ connecting them). We denote by $\overline{G}$ the complement graph, which has the same vertex set, and the complementary set $\overline{E} \in V\times V$ of edges. We will occasionally be somewhat sloppy in notation and write $e\in G$ ($e\not\in G$)to indicate that $e$ is (not) an edge in $G$ (and thus $e\in E$ ($e\not\in E$)). Similarly we will sometimes write subsets of edges as $S \subseteq G$. A subgraph $H$ of $G$ is a obtained from $G$ by considering a subset of the edges $E'\subseteq E$ and an induced subgraph $H$ is obtained by considering a subset of the \emph{vertices} $V'\subseteq V$, which has edges $E' = E\cup V'\times V'$.\\

The symbol $G_{n, d}$ refers to an $n$-vertex $d$-regular graph and $\overline{G}_{n,d}$ refers to the complement of the same graph . We drop the subscripts when the number of vertices and the regularity parameter are clear from the context. For a graph $G$, define $m(G)$, the density of the graph, as
\begin{equation}
m(G) = \text{max}\bigg\{\frac{|E|_{H'}}{|V|_{H'}},~ H' \subseteq G, ~~|V|_{H'} > 0 \bigg\},
\end{equation}
An $n$-qubit graph state $\ket{G}$ is defined in terms of an $n$-vertex graph $G$ as
\begin{equation}
\ket{G}  =  \prod_{(i,j): U_G[i,j] = 1} CZ_{i,j} \ket{+^n} = \frac 1 {\sqrt{2^{n}}} \sum_{x\in \{0,1\}^n} (-1)^{x^TU_G x} \ket{x},
\end{equation}
where $U_G$ is the upper triangle of the adjacency matrix $A_G$ of $G$ and the inner product is taken over $\mathbb{F}_2$. Graph states are a type of stabilizer state (which we will denote by $\ket{S}$), which means they are the joint $+1$ eigenvectors of a set of $2^n$ mutually commuting Pauli matrices. There are $2^{n(n-1)/2}$ graph states and $2^n \prod_{i=1}^n (2^i+1)$ stabilizer states. A key property of stabilizer states is the following expectation value (taken uniformly over the stabilizer states), first derived in \cite{gross_schurweyl_2021}:
\begin{equation}\label{eq:average_stab}
\mathbb{E}_{\ket{S}} \ket{S}\!\bra{S}\tn{t} = \frac{1}{2^n\prod_{i=0}^{t-2} (2^n+ 2^i)} \sum_{T\in \Sigma_{t,t}}R(T),
\end{equation}
where $\Sigma_{t,t}$ is the set of Lagrangian subspaces of $\mathbb{F}^{2t}_2$ with respect to a particular generalized quadratic form, and $R(T)$ is an $nt$-qubit representation of this set (which can be given a semigroup structure). We will need very few properties of this set (see \cite{gross_schurweyl_2021} for an exhaustive treatment), only that
\begin{align}
|\Sigma_{t,t}| = \prod_{i=0}^{t-1}(2^i+1),\\
\tr (\ket{\beta}\!\bra{\beta}\tn{t} R(T)) \leq 1,
\end{align}
for all states $\ket{\beta}$. The latter statement follows by combining Proposition $56$ and Theorem $72$ in \cite{bittel2025complete} (a more direct statement, using a different proof technique, can also be found as Corollary $6.11$ in \cite{hinsche2025clifford}).

\subsection{Random graph models}\label{subsec:models}
We will consider three different models of random regular graphs (with regularity parameter $d$), each with slightly different properties. 
\begin{definition}[uniformly random $d$-regular graphs]
$G(n,d)$ is the distribution over graphs generated by choosing $d$-regular graphs uniformly at random.
\end{definition}
\begin{definition}[Matching model of $d$-regular multigraphs]
$G_m(d,n)$ for $n$ even is the distribution over regular multigraphs of degree $d$ generated by choosing $d$ matchings independently uniformly at random and composing the result.
\end{definition}
\begin{definition}[Pairing model of $d$-regular multigraphs]
$G_p(d,n)$ for $dn$ even is the distribution over regular multigraphs of degree $d$ generated by choosing a uniformly random matching $M$ on the set $\{(i,j)\;\|\; i\in [n], j\in [d]\}$ and adding an edge between $i,\hat{i}$ whenever $(i,j),(\hat{i},\hat{j})$ is an edge in $M$.
\end{definition}
All three of these models can be efficiently sampled from even when $d$ is relatively large~\cite{gao2017uniform}. Note that the pairing model allows for both multiple edges between vertices and self-edges (loops).
For $d$ constant the pairing model and uniformly random graph model are related. The following lemmas are well known in the graph theory literature~\cite{wormald1984generating}:
\begin{lemma}
The probability that the pairing model yields a simple $d$-regular graph is bounded from below as
\begin{equation}
\mathbb{P}\big[G\; \text{is simple}\big] \geq 2^{-d^2}.
\end{equation}
\end{lemma}
\begin{lemma}
Conditioned on being simple, the graphs obtained from the configuration model are \emph{uniformly distributed} $d$ regular graphs.
\end{lemma}
We have a similar simplicity condition in the matching model.
\begin{lemma}
The probability that the random matching model yields a simple $d$-regular graph is bounded from below as
\begin{equation}
\mathbb{P}\big[G\; \text{is simple}\big] \geq 2^{-d}.
\end{equation}
\end{lemma}
However in this model, conditioning on simplicity does not yield a uniform distribution on $d$-regular graphs. This can be easily seen by noting that there exist $d$-regular graphs that do not contain a perfect matching. The conditional distribution is equivalent to uniformly random $d$-regular graphs in a weaker sense, called \emph{contiguity} \cite{janson_random_1995}, meaning that event that happen with high probability in one distribution also happen with high probability in the other. We will not make use of this connection in this paper and consider the matching model to be interesting in its own right. 

\subsection{Krawtchouck polynomials}

\label{ssec:krawtchouk prelims}

Krawtchouk polynomials are a family of polynomials that prominently appear in classical error correction codes and boolean analysis. We will need them in the proof of \Cref{thm:anticoncentration random pairing}. We will briefly recap their definition here as well as a number of upper bounds. The (binary) Krawtchouk polynomial of degree $i$ and size $N$ is defined as
\begin{equation}
\label{eq:krawtchouk}
K_i^N(x) \coloneqq \sum_{q=0}^i(-1)^q \binom{x}{q}\binom{N-x}{i-q}.
\end{equation}
These polynomials are orthogonal under a binomial measure (see e.g. \cite{feinsilver2005krawtchouk}), i.e.
\begin{equation}
\sum_{t = 0}^N \binom{N}{t}K_i^N(t) K_j^N(t) = 2^N\binom{N}{i}\delta_{ij}.
\end{equation}
This immediately implies a pointwise upper bound for integer evaluations of the polynomial of the form
\begin{equation}\label{eq:upper_bound_orth}
|K_i^N(t)|\leq 2^{N/2} \binom{N}{i}^{1/2} \binom{N}{t}^{-1/2}.
\end{equation}
This is a rather straightforward upper bound but it will service almost all of our needs (it is also surprisingly close to being tight, see \cite{kirshner2021moment}). However to cover certain parameter regimes we will also need a more sophisticated bound from \cite[Corollary 16]{derksen_pseudorandomness_2024} of the form
\begin{equation}\label{eq:upper_bound_derksen}
|K_i^N(t)|\leq \binom{N}{i} \bigg(\frac{i}{N} + \frac{(N-t)^2}{N^2}\bigg)^{i/2}.
\end{equation}

\subsection{Drift and minorization of Markov chains}

\label{ssec:markov prelims}

Consider a Markov chain $P$ on a (possibly countably infinite) state space $\mc{X}$. We will maintain that a Markov matrix acts from the left, that is $P(x,x')\coloneqq \mathbb{P}(X_{t+1}= x'|X_{t} = x)$ for $x, x'\in \mc{X}$. If the Markov chain is irreducible it will have a unique stationary distribution $\pi$. Bounding the convergence time of the Markov chain towards $\pi$ on unbounded systems is generally tricky. In this paper we will use the drift and minorization method \cite{meyn1994computable}, which can bound convergence from a fixed starting state to the stationary distribution for an irreducible Markov chain in a way that does not depend directly on the size of $\mc{X}$. This method consists of two steps, first bounding the time it takes for the Markov chain to pool into a ``small set'', and then bounding thermalization within that set. The first requirement (the ``drift'') is the existence of a \emph{drift} function $V$ which controls the convergence to a \emph{small set} $C$:
\begin{definition}[drift towards a small set]
A function $V:\mc{X}\to [0, \infty] $ is a drift function (towards a set $C\subset \mc{X}$) for a Markov chain $P$ if there exists a constant $0<\lambda<1$ and a constant $b<\infty$ such that
\begin{equation}
P V(x) \leq \lambda V(x) + b I_C,
\end{equation}
where $I_c$ is the indicator function on the set $C$ and $P V(x) \coloneqq \sum_{x'\in \mc{X}} P(x,x')V(x'))$.
\end{definition}

The second requirement is a minorization condition on the small set $C$
\begin{definition}[minorization on small set]
A Markov chain $P$ satisfies a minorization condition on the set $C$ if there exists a probability distribution $\nu$ on $C$ and a constant $\delta>0$ such that for all $x \in C$ we have 
\begin{equation}
P(x,x') \geq \delta \nu(x').
\end{equation}
\end{definition}
Finally, the drift and minorization conditions are called \emph{compatible} if there exists a constant $d > 2b/(1-\lambda)$ such that the level set $\{x\in \mc{X}\;\|\; V(x)\leq d\}$ is included in the small set $C$. 

If a Markov chain has a drift function and satisfies the minorization condition on a small set $C$ in a compatible way, we can bound the convergence of the Markov chain to its stationary distribution in an exponential way. We will use the following theorem due to Rosenthal \cite[Thm.12]{rosenthal1995minorization}:
\begin{theorem}\label{thm:markov_conv}
Let $P$ be a Markov chain on a (countable) state space $\mc{X}$, compatibly satisfying a drift and minorization condition with function $V(x)$, small set $C$ and parameters $\lambda, b, \delta, d$. Let $\nu$ be an initial distribution and let $\pi$ be the stationary distribution of $P$. We now have for all $0<r<1$ that:
\begin{equation}
\norm{P^k\nu -\pi}_{TV} \leq (1-\delta)^{rk} + (1+ 2b/(1-\lambda) + \mathbb{E}_\nu(V))\bigg[\bigg(\frac{1+ 2b +\lambda d}{1+d}\bigg)^{1-r} \big(1+ 2\lambda d+ b\big)^r\bigg]^k,
\end{equation}
where $\norm{\cdot}_{TV}$ is the total variation distance.
\end{theorem}
Due to the many free parameters this theorem is quite flexible, but also rather difficult to use (and gives quite conservative bounds). We will use it in the proof of \Cref{thm:geom_ent_upper}.

\section{Anticoncentration of graph states of constant degree}\label{sec:constant}

In this section we will compute the average (normalized) second moment  for two ensembles of random graphs with degree $d$ which induce measures on $d$-regular graphs.

First, we compute the average second moment for a random multi-graph in the random pairing model. Conditioning on simplicity will then give results for uniformly random $d$-regular (simple) graphs, as discussed in \Cref{subsec:models}. 
To obtain the result in \Cref{thm:anticoncentration} for  the random matching model, we will just need to slightly adapt the proof of this more complicated case.

\begin{theorem}[Anticoncentration of random pairing model graph states (Restatement of part (a) of \Cref{thm:anticoncentration})]
\label{thm:anticoncentration random pairing} 
Consider the uniform measure $\mc G_p(n,d)$ on $d$-regular multi-graphs chosen from the pairing model.
Then, for any  $2< d =o(n^{1/2})$ we have
\begin{equation}
\mathbb{E}_{G \sim \mc G_P(n,d),\theta}(m_2(G, \theta))   =  o(1) + \begin{cases}
    2 & \text{ if } d = 1 \mod 2\\
    3 & \text{ if } d = 0 \mod 2 .
\end{cases}
\end{equation}
\end{theorem}

\begin{corollary}[Restatement of \Cref{thm:random regular}]
Consider the uniform measure $\mc G_r(d,n)$ on $d$-regular graphs on $n$ vertices, and uniformly random angles $\theta$. 
Then for any constant $d$
\begin{equation}
\mb E_{G \sim \mc G_r(d,n), \theta} [m_2(G,\theta)]  \leq (3+o(1)) 2^{d^2}.
\end{equation}
\end{corollary}
\begin{proof}
    The corollary follows directly from \Cref{thm:anticoncentration random pairing}, observing that the probability that a random multi-graph in the pairing model $G \leftarrow \mc G_p(d,n)$ is simple, is lower bounded by $1/2^{d^2}$, and hence, conditioning on the event that a graph in $\mc G_p(d,n)$ is simple yields the statement.
\end{proof}

\subsection{Anticoncentration of a fixed graph state}
The proof of \Cref{thm:anticoncentration random pairing} is based on a series of lemmata. 
As a first step, we find an expression for the second moment of a fixed graph states, averaged over the random choice of measurement angles. 
This follows more or less directly from the discussion in Appendix E of \cite{hangleiter_fault-tolerant_2025}.   
\begin{lemma}\label{lem:graph_xeb}
Consider a graph $G$ and a uniformly random choice of $X$-$Y$-plane measurement angles $\theta$. 
Then we have that 
\begin{equation}
\label{eq:second moment angle average}
\mathbb{E}_\theta(m_2(G, \theta))  = 2^{-n}\sum_{L,R\subset [n], L\cap R = \emptyset} (-1)^{|A_G[L,R]|},
\end{equation}
where $A_G$ is the adjacency matrix of $G$. 
$|A_G[L,R]|$ denotes the sum of the entries of the submatrix $A_G[L,R]$ of $A_G$ corresponding to rows in $L$ and columns in $R$, and counts the number of edges crossing between $L$ and $R$. 
\end{lemma}
\begin{proof}
To show the lemma, we utilize the statistical-mechanics mapping of second moments of IQP circuits derived in Ref.~\cite{hangleiter_fault-tolerant_2025}, extended to arbitrary single-qubit $Z$ rotations.
We keep the discussion of this model brief here and refer the reader to Appendix E of Ref.~\cite{hangleiter_fault-tolerant_2025} for the derivation of the statistical-mechanics mapping.
We start from observing that the second moment operator 
\begin{align}
    M_2 = \mb E_\psi \proj \psi \otimes \proj \psi,
\end{align}
of a ensemble of random states $\ket \psi$ determines the second moment of its outcome distribution $p_\psi$ as $\mb E_\psi p_\psi(x)^2 = \bra x^{\otimes 2} M_2 \ket x^{\otimes 2}$, and thus computing the average second moment can be reduced to computing the second moment operator of the underlying state ensemble. For a random product rotation around the $Z$ axis applied to the $\ket +^{\otimes n}$ state the second moments given by 
\begin{align}
    \mb E_\theta\left[ \ee^{-i \sum_j \theta_j Z_j} \proj{+^n} \ee^{i \sum_j \theta_j Z_j} \right] = \frac {1 }{4^n} ( \ix + \sx + \xx)^{\otimes n} = \frac 1 {4^n}\sum_{Q \in \mc S^{\otimes n}} Q ,
\end{align}
where $\ix = \proj{01} + \proj{10}$, $\sx = \ket{01}\bra{10} + \ket{10}\bra{01}$, $\xx = \proj {00} + \proj{11}$ and we let $\mc S = \{\ix, \sx, \xx\}$ be the set of possible `states' of the statmech model.
Next, we observe that 
\begin{align}
    CZ^{\otimes 2} (P \otimes Q) CZ^{\otimes 2} = \begin{cases}
        - P \otimes Q & \text{ if } P=\ix, Q = \sx \wedge P=\sx, Q = \ix,\\
        P \otimes Q & \text{ else,} 
    \end{cases}
\end{align}
and that $
    \sum_{x \in \bin}\bra {xx}H^{\otimes 2} Q H^{\otimes 2} \ket {xx} =  1, \,  \forall Q = \ix, \sx, \xx. $
Hence, the sign of a particular `state' $Q \in \mc S^{\otimes n}$ on a graph $\ket G$ is given by the parity of the $CZ$ gates that act on an $\ix \sx$ or $\sx \ix$ pair.

Thus, the second moment of the outcome distribution $p_{G, \theta}$ of measuring a graph state $\ket G$ in the $X$-$Y$ plane angles $\theta$, averaged over the random choice of $\theta$, can be written as
\begin{align}
    \mb E_\theta[ m_2(G, \theta) ] %
 = \frac 1 {2^n} \sum_{Q \in \mc S^{\otimes n}} (-1)^{N(Q,G)},
 \end{align}
where for $G = (V,E)$ with edge set $E$ we let $N(Q,G) = |\{ (e_0, e_1) \in E: Q_{e_0}\otimes  Q_{e_1}\in \{ \ix\otimes \sx, \sx \otimes \ix \}\}|$ be the number of edges in the graph $G$ coinciding with a $\ix\sx$ or $\sx \ix$ pair in the state $Q$. 
Since this expression only depends on the locations of the graph edges and $\ix$ and $\sx$ states, we can rewrite it in the form \eqref{eq:second moment angle average}.
\end{proof}

Importantly, this expression can also be interpreted for multi-graphs, where the elements of the adjacency matrix are now integers.
The crucial number $|A_G[L,R]|$ featuring in \Cref{eq:second moment angle average} thus still counts the number of edges crossing between the sets $L$ and $R$. 

\subsection{Averaging over matchings}
Next, we need to evaluate the average over the random choice of (multi-)graphs. 
Since our graphs are defined in terms of uniformly random matchings, we begin by analyzing the average parity of edges in a random matching crossing between two disjoint sets $L,R \subset [n]$.
We first note some symmetry properties of the average parity in \Cref{lem:symmetry}, and then express the average parity in terms of Krawtchouk polynonmials in \Cref{lem:krawtchouk}.
In the following, let $\mc M(n)$ denotes the uniform distribution over perfect matchings of $n$ vertices.

\begin{lemma}\label{lem:symmetry}
Consider a random matching $M$ on $n$ vertices and three disjoint sets $L,R,T\subseteq[n]$ covering   $[n] = R \cup L \cup T$. The value of the average parity of the matching between two sets is invariant under the interchange of $R,L,T$ up to factors of $\pm 1$ as 
\begin{align}
\label{eq:exchange L R}
\mathbb{E}_{M \sim \mc M(n)} (-1)^{|A_{M}(L,R)|} &= \mathbb{E}_{M  \sim \mc M(n)} (-1)^{|A_{M}(R,L)|}\\
\label{eq:replace R T}
\mathbb{E}_{M \sim \mc M(n)} (-1)^{|A_{M}(L,R)|} &= (-1)^{|L|}\mathbb{E}_{M  \sim \mc M(n)} (-1)^{|A_{M}(L,T)|}. 
\end{align}
\end{lemma}
\begin{proof}
Clearly the average parity is invariant under the interchange of the arguments, i.e.,   
\begin{align}
\mathbb{E}_{M} (-1)^{|A_{M}(L,R)|} &= \mathbb{E}_{M} (-1)^{|A_{M}(R,L)|},
\end{align}
To see \Cref{eq:replace R T},
consider a particular matching that has $q$ edges going from $L$ to $R$ (and thus contributes $(-1)^q$ to the average). This matching also has $0 \le q' \le n - q $ edges going from $L$ to $T$. By construction the remaining vertices in $L$ must be matched to each other. Hence $q +q'= |L| \mod 2$ and therefore $(-1)^q = (-1)^{q'} (-1)^{|L|}$. Since this is true for every matching the claim follows.
\end{proof}

\begin{lemma}\label{lem:krawtchouk}
Consider two disjoint sets $L,R\subseteq [n]$ such that $|L|,|R|\leq n-|L|-|R|$. Then the average parity of the number of edges of a random matching between $L$ and $R$ is given by
\begin{equation}
\mathbb{E}_{M \sim \mc M(n)} (-1)^{|A_{M}(L,R)|} = \sum_{\substack{i=0\\i=|L| \mod 2}}^{|L|} K^{n-|L|}_i(|R|) \frac{|L|! (n-|L| - i-1)!!}{(|L|-i)!! (n-1)!!},
\end{equation}
where $K_i$ is the Krawtchouk polynomial of degree $i$.
\end{lemma}
\begin{proof}
Denote $T = [n]\backslash (R\cup L)$. Consider a matching that has $q$ edges between the sets $L$ and $R$ and $q'$ edges between $L$ and $T$. Note that this is only possible if $q+ q' = |L| \mod 2$ and $q+q'\leq |L|$. There are $\binom{|L|}{q}\binom{|R|}{q} \cdot q!$ ways to choose the first $q$ edges and $\binom{|L|-q}{q'}\binom{|T|}{q'} \cdot q'!$ ways to choose the second set of edges. Once these edges are fixed there are $(|L| - q-q'-1)!! (n- |L| - q-q'-1)!!$ ways to complete this edge set to a full matching of $[n]$. Hence we can write
\begin{multline}
\mathbb{E}_{M} (-1)^{|A_{M}(L,R)|} = \frac {1}{(n-1)!!} \hspace{-2em} \sum_{\substack{q,q'\geq 0\\q+q'\leq |L|\\q+ q' = |L| \mod 2\\ q\leq |R|}}\hspace{-2em}(-1)^q q!\binom{|L|}{q}\binom{|R|}{q}q'!\binom{|L|-q}{q'}\binom{|T|}{q'} \\ \times(|L| - q-q'-1)!! (n- |L| - q-q'-1)!!,
\end{multline}
since there are $(n-1)!!$ matchings in total. Since $\binom{|R|}{q}=0$ if $|R|<q$ we can drop the $ q\leq |R|$ constraint going forward.
We can expand the binomials involving $|L|$ and use the identity $(a-1)!!\, a!! = a!$ to rewrite this as 
\begin{equation}
\mathbb{E}_{M} (-1)^{|A_{M}(L,R)|} = \sum_{\substack{q,q'\geq 0\\q+q'\leq |L|\\q+ q' = |L| \mod 2}} (-1)^q  \binom{|R|}{q} \binom{|T|}{q'}  \frac{|L|! (n- |L| - q-q'-1)!!}{(|L|-q-q')!!(n-1)!!}.
\end{equation}
Changing variables $q,q' \rightarrow q, i = q+q'$ we can rewrite this further as
\begin{equation}
\mathbb{E}_{M} (-1)^{|A_{M}(L,R)|} = \sum_{\substack{i\geq 0\\ i = |L| \mod 2}}^{|L|}\sum_{q = 0}^{i} (-1)^q  \binom{|R|}{q} \binom{|T|}{i-q}  \frac{|L|! (n- |L| - i-1)!!}{(|L|-i)!!(n-1)!!}.
\end{equation}
in which we recognize the definition \eqref{eq:krawtchouk} of the Krawtchouck polynomial.
\end{proof}

\subsection{Proof of \texorpdfstring{\Cref{thm:anticoncentration random pairing}}{}}
We can now provide a proof of the core result of this section (\Cref{thm:anticoncentration random pairing}). This proof is long, but much of the conceptual has been done in the previous subsections (with `merely' some counting remaining).
\begin{proof}[Proof of \Cref{thm:anticoncentration random pairing}]
\label{proof:anticoncentration pairing}
\Cref{lem:symmetry,lem:krawtchouk} imply that all that is relevant for the average parity of the edges between two subsets $L, R \subseteq [n]$ is the size of the subsets. 
Recall that these sets correspond to $\ix$ and $\sx$ states on a subset of the vertices and let us think of the different states on the vertices as one of three colors ($L\leftrightarrow \ix$, $R \leftrightarrow \sx$, $T \leftrightarrow \xx$). 
To analyze the random pairing model,  we observe the following: 
for a fixed choice of $L,R \subseteq [n]$ in the sum \eqref{eq:second moment angle average}, the average parity $\mb E_{{G_M}} [(-1)^{|A_{G_M}[L,R]|}]$ over graphs $G_M$ induced by a matching $M$ (the number of edges between $\ix$-colored vertices $\sx$-colored vertices) is equal to the average parity $\mb E_M[(-1)^{|A_M[L_d,R_d]}]$ of two sets $L_d, R_d \subseteq [dn]$ of size $|L_d| = |L|d, |R_d| = |R|d$ over the inducing matching. 
This is because we can `split up' every vertex in $G_M$ into $d$ vertices of the same color.
All that matters for the average parity is the crossings between colors blowing up the sets $L \rightarrow L_d, R \rightarrow R_d$. 

Hence, it follows from \Cref{lem:graph_xeb,lem:symmetry,lem:krawtchouk} that 
\begin{align}
\mb E_{G_M\sim \mc G_p(n,d),\theta}(m_2(G_M, \theta) &= 2^{-n}\sum_{L,R\subset [n], L\cap R = \emptyset} \mathbb{E}_{G_M\sim \mc G_p(n,d)}(-1)^{|A_{G_M}[L,R]|} \\
&= 2^{-n}\sum_{L,R\subset [n], L\cap R  = \emptyset} \mathbb{E}_{M \sim \mc M(nd)}(-1)^{|A_M[L_d,R_d]|} .
\label{eq:division L R average}
\end{align} 
Let us denote (counter-intuitively, we are following the notation in \cite{hangleiter_fault-tolerant_2025}) $|L| = k, |R| = l$ (correspondingly $|L_d| = dk, |R_d| = dl$). Then we can use \Cref{lem:krawtchouk} to get 
\begin{align}
\label{eq:second moment krawtchouk k}
\mb E&_{G_M\sim \mc G_p(n,d),\theta} \left[m_2(G_M, \theta)\right] =2^{-n} \sum_{\substack{L,R\subset [n]\\ L\cap R  = \emptyset}} \sum_{\substack{i=0\\i=dk \mod 2}}^{dk} K^{d(n-k)}_i(dl) \frac{(dk)! (d(n-k) - i-1)!!}{(dk-i)!! (dn-1)!!} 
\\
 &=2^{-n} \sum_{k=0}^n \sum_{l=0}^{n-k} \binom n k \binom {n-k} l \underbrace{\left[\sum_{\substack{i=0\\i=dk \mod 2}}^{dk}K^{d(n-k)}_i(dl) \frac{(dk)! (d(n-k) - i-1)!!}{(dk-i)!! (dn-1)!!}  \right]}_{\eqqcolon M_p(n,k,l,d)},
\end{align}
where the second equality counts the number of possible choices of $L,R \subseteq [n]$ with fixed sizes $k,l$ and in the last equality we have implicitly defined $ M_p(n,k,l,d)$. 
To get a feeling for what we need to show, consider \Cref{fig:ring plot}. There, we show the term $ M_p(n,k,l,d)$ for fixed $n,d$ as a function of $k,l$. There is a clear separation of what is happening. 
At the boundaries, i.e., for $k=0$ or $l=0$ or $k=l$ the terms are largest, and they decay towards the center. 
\begin{figure}
    \centering
    \includegraphics{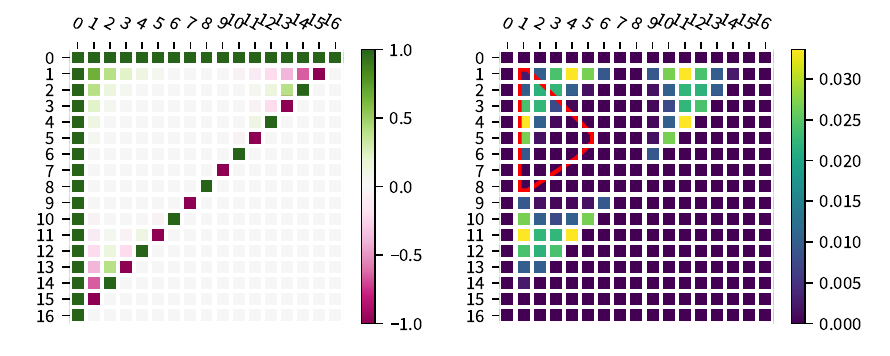}
    \caption{\label{fig:ring plot} For $n = 16,d = 3$, we show the dependence on $k,l$ of \emph{(left)} $M_p(n,k,l,d)$ and \emph{(right)}  $2^{-n } \binom nk \binom{n-k}{k} |M_p(n,k,l,d)|$ for $k,l > 0 $ and $k+l<n$. The red area corresponds to the region $1\le k \le l \le n -k-l$ in which we need to achieve a nontrivial bound.}        
\end{figure}

This suggests to divide up the sum according to this behavior. 
We can also see the symmetry under the exchange $k \leftrightarrow l$, which we will exploit later. 
Specifically, we will divide the sum over $k,l$ into four  cases. The only non-negligible one will be  Case 0 ( corresponding to $k=0$, $l=0$, or $k+l =n$). 

\paragraph{Case 0:}
Let us begin by considering the case in which one of $L,R,T$ is the empty set, i.e., $k=0 \vee l= 0 \vee n-k-l = 0$. We begin by treating the $k=0$ sub-case (we label the associated term in the overall sum $T_{k=0}$), the others will follow by symmetry:
\begin{align}
2^n & T_{k=0} = \sum_{R\subset [n]} \mathbb{E}_{M \sim \mc M(nd)}(-1)^{|A_{M}[\emptyset,R_d]|}\nonumber \\ & \qquad + \sum_{L\subset [n]} \mathbb{E}_{M \sim \mc M(nd)}(-1)^{|A_{M}[L_d,\emptyset]|}  + \sum_{\substack{L, R\subset [n]\\ L \cap R = \emptyset, L \cup R = [n]}} \mathbb{E}_{M \sim \mc M(nd)}(-1)^{|A_{M}[L_d,R_d]|} \\
&= \sum_{R\subset [n]} \left( 2 \cdot \mathbb{E}_{M \sim \mc M(nd)}(-1)^{|A_{M}[\emptyset,R_d]|}  +  (-1)^{|R_d|} \mathbb{E}_{M \sim \mc M(nd)}(-1)^{|A_{M}[\emptyset,R_d]|}  \right).
\end{align}
Now, observe that $|R_d| = ld$ and therefore $(-1)^{ld} =1 $ for even $d$ and $(-1)^{ld} = (-1)^l$ for odd $d$. 
By \Cref{lem:symmetry} $\mathbb{E}_{G_M\sim \mc G_p(n,d)}(-1)^{|A_{G_M}[\emptyset,R]|} = \mathbb{E}_{G_M\sim \mc G_p(n,d)}(-1)^{|A_{G_M}[\emptyset,[n]\setminus R]|}$ the second term thus vanishes for odd $d$ and is equal to the first term for even $d$.

We can now simplify the remaining sum to $|L| \le n- |L|$ using \Cref{lem:symmetry} incurring a factor of $2$ and counting the number of choices of $L \subset n$
\begin{align}
{\mathrm{Case}\;0} =2^{-n} \sum_{l \le n-l} \binom n l \mathbb{E}_{M \sim \mc M(nd)}(-1)^{|A_{M}[\emptyset,R_d]|}  \cdot \begin{cases}
    4 & \text{ if } d = 1 \mod 2\\
    6 & \text{ if } d = 0 \mod 2 .
\end{cases}
\end{align}
Furthermore, we can use \Cref{lem:krawtchouk}to get 
\begin{align}
 \mathbb{E}_{M \sim \mc M(nd)}(-1)^{|A_{M}[\emptyset,R_d]|} =   K^{dn}_0(dl) \frac{(dn-1)!!}{(dn-1)!!},
 \end{align} 
 and the fact that  $K^{dn}_0(dl) =1$ (see \Cref{eq:krawtchouk}) to get 
\begin{equation}
{\mathrm{Case}\;0} = 2^{-n} \sum_{0\leq l \leq n }\binom{n}{l} \cdot \begin{cases}
        4/2 & \text{ if } d = 1 \mod 2\\
    6/2 & \text{ if } d = 0 \mod 2 .
\end{cases} \\
= \begin{cases}
        2 & \text{ if } d = 1 \mod 2\\
   3 & \text{ if } d = 0 \mod 2 .
\end{cases}
\end{equation}
It remains to bound the remaining terms. 
We begin by defining the sets
\begin{align}
W_n^* & \coloneqq \{(k,l) \in [n]^{\times 2}: 0 < k \le l \le n - k-l \}\\
\label{eq:def S}
\mc S & \coloneqq \bigg\{(k,l)\in W^*_n \;\bigg | \;\;2^{-n} \binom{n}{k} \binom{n-k}{l} \geq n^{-3}\bigg\}.
\end{align}
This set captures the terms in the sum where we need to achieve a nontrivial bound. 
This is because the sum over $k,l$ runs over at most $n^2$ terms and the average over the parity (the remaining part of \Cref{eq:expression for average m2}) is bounded by $1$ so that the sum over all $(k,l) \in W_n^* \setminus \mc S$ is bounded by $O(1/n)$. 
Using \Cref{lem:symmetry}, we can always assume that $k\leq l\leq n-k -l$ (this gives a factor of $3! = 6$ in the expression) and thus compute the remaining $k\ge 1$ terms of \Cref{eq:second moment krawtchouk k} as 
\begin{align}
\label{eq:expression for average m2}
T_{k \ge 1}  &=   6  \!\! \!\! \sum_{1\leq k\leq l \leq n-k-l } 2^{-n} \binom{n}{k}\binom{n-k}{l} M_p(n,k,l,d)\\
&= O(1/n) + 6  \!\! \!\! \sum_{(k,l) \in \mc S} 2^{-n} \binom{n}{k}\binom{n-k}{l} M_p(n,k,l,d)\\.
\end{align}
We will use different bounds in different regimes of the sum over $k,l \in \mc S$, decomposing $T_{k \geq 1} =  O(1/n) + T_{\text{Case 1}} + T_{\text{Case 2}} + T_{\text{Case 3}}$ according to the following three remaining cases. 
\begin{description}
    \item[Case 1:] $1 \le k \leq 90 \log n$,
    \item[Case 2:] $k \geq 90 \log n$ and $2^{-n} \binom{n}{k} \binom{n-k}{l} \geq n^{8}$,
    \item[Case 3:] $k \geq 90 \log n$ and $n^{-3}\leq 2^{-n} \binom{n}{k} \binom{n-k}{l} \leq n^{8}$.
\end{description}
We will now treat each of these cases in turn.
\paragraph{Case 1 ($1 \le  k < 90 \log n $):}
For the first case, we will use the upper bound for Krawtchouck polynomials due to Derksen, given in \Cref{eq:upper_bound_derksen}, in order to  bound $M_p(n,k,l,d)$ 
\begin{multline}
\label{eq:derksen bound T_1 le k le 8 log n}
M_p(n,k,l,d) = \sum_{\substack{i=0\\i=dk \mod 2}}^{dk}  K^{d(n-k)}_i(dl) \frac{dk! (d(n-k) - i-1)!!}{(dk-i)!! (dn-1)!!}\\
\leq\sum_{\substack{i=0\\i = dk \mod 2}}^{dk} \binom{d(n-k)}{i}\bigg(\frac{i}{d(n-k)} + \frac{(n-k-2l)^2}{(n-k)^2}\bigg)^{i/2}\frac{(dk)!(d(n-k) - i - 1)!!}{(dk-i)!!(dn-1)!!}.
\end{multline}

To further bound this, we consider the factors  of \Cref{eq:derksen bound T_1 le k le 8 log n} individually 
\begin{description}
    \item[Factor 1] For the first factor, we have the crude bound. 
\begin{equation}
\binom{d(n-k)}{i}\leq d^i(n-k)^i .
\end{equation}
\item[Factor 2]
To bound the second factor, we need a lower bound on $l$. We get this by noting that $k,l\in \mc S$ implies that
\begin{equation}
2^{(n-k)H(l/(n-k))}\geq \binom{n-k}{l}\geq \frac{2^n}{n^3 \binom{n}{k}}\geq 2^n n^{-3-k}.
\end{equation}
Taking the log of both sides and using $H(x) \leq 2 \sqrt{x(1-x)}$, we obtain
\begin{equation}
\frac{l}{n-k} \bigg(1-\frac{l}{n-k}\bigg) \geq \bigg(\frac{n -(3+k)\log(n)}{2(n-k)}\bigg)^2.
\end{equation}

Solving the above quadratic inequality, using that $k \le 90 \log n$, we get the bound
\begin{equation}
 \frac{n-k}{2} - C_0 \sqrt n \log n \le l \le \frac {n-k} 2 , 
\end{equation}
for some constant $C_0$ and all $n$ larger than some constant $n_0$. 
Since $l\leq n/2$ we thus have that
\begin{align}
    \left( \frac{n-k-2l}{n-k}\right)^2 \le 
    \left( 1 - \frac 1 2 + \frac{C_0 \sqrt n \log n}{n-k}\right)^2 \le     \left( \frac 1 2 + \frac{C_0 \log n}{\sqrt n-90 \log n /\sqrt n}\right)^2 \le C_1 \log ^2 (n), \notag
\end{align}
for all $n \ge n_0$ and some constant $C_1$.
Moreover, 
\begin{equation}
\frac{i}{d(n-k)} \leq \frac{90\log(n)}{n-90\log{n}}\leq 1 \leq C_2\log^2(n).
\end{equation}
Altogether, we thus get for all $n \ge n_1 = const.$ 
\begin{align}
    \frac{i}{d (n-k)} + \left(\frac{n-k-l}{n-k} \right)^2 \le C_2 \log ^2(n),
\end{align}
with constant $C_2$.
\item[Factor 3] For the last factor we have the bound 
\begin{align}
\frac{(dk)! (d(n-k) - i - 1)!!}{(dk-i)!!(dn-1)!!} &\leq (dk)! 
(d(n-k)-i-1)^{-(dk+i)/2} \\
&\leq (dk)!(dn-2dk-1)^{-(dk+i)/2}. 
\end{align}
\end{description}
\noindent
Putting the bounds for Factors 1-3 together, we thus find
\begin{align}
M_p(n,k,l,d) &\le \sum_{\substack{i=0\\i = dk \mod 2}}^{dk}  d^{(i-dk)/2} (n-k)^i  (n-2k-1/d)^{-(dk+i)/2}(dk)!(C_2 \log n)^i \\
&\le \sum_{\substack{i=0\\i = dk \mod 2}}^{dk}  (n-k)^{i/2}
  (n-2k-1)^{-dk/2}
   (dk)!  (C_2\log n)^i.
\end{align}
The dominant term in the sum is the $i=dk$ term for which we have 
\begin{align}
    \left(\frac{n-k}{n-2k-1}\right)^{dk/2} = \left(\frac{1}{1-\frac{k-1}{n-k}}\right)^{dk/2} \in O(n)^{dk/2}.
\end{align}
This gives us
\begin{align}
M_p(n,k,l,d)   \le  dk\, (dk)!C_3^{dk}  n^{-dk/2} \log(n)^{dk},
\end{align}
for some constant $C_3$.
Now we can go back to upper bounding the full expression \eqref{eq:expression for average m2}. 
Since $l\leq (n-k)/2$ by assumption we also note that the set $S_k = \{l\; | \;(k,l) \in \mc S\}$ has size $|S_k|\leq \sqrt{n}\log(n)C_0 $. 
Moreover, we observe that 
\begin{align}
    \binom n k \binom{n-k} l \le n^k 2^{n-k}\frac{\sqrt 8}{\sqrt{n-k}},
\end{align}
and thus obtain with a constant $C_4$
\begin{multline}
2^n|T_{\mathrm{Case}\;1}| \leq 6 \sum_{k=1}^{90\log(n)} 2^{n-k}\frac{ n^{k+1/2}}{\sqrt{n-k}} dk (dk)! C_4^{dk}\log(n)^{dk+1} n^{-dk/2}\\
\le6 \sum_{k=1}^{90\log(n)} 2^{n-k}\frac{ n^{k+1/2}}{\sqrt{n-k}}  (C_4)^{dk}\log(n)^{dk+1}(dk)^{dk+1} n^{-dk/2}.
\end{multline}
Since $k^{d+1}\cdot90^{d+1}\log(n)^{d+1}\leq n^{d/2}$ for $d>2$ and sufficiently large $n$, the terms in the sum are strictly decreasing in $k$. 
Hence, the sum is dominated by the $k=1$ term and we get for any
 $2 < d = o(n^{1/2})$
\begin{equation}
|T_{\mathrm{Case}\;1}| \in O\left( d^{d+1} \log(n)^{d+3} n^{-d/2+1}  \right) = o(1),
\end{equation}
which is all we need. \\


\paragraph{Case 2 ($k \ge 90 \log n$ and $2^{-n} \binom{n}{k} \binom{n-k}{l} \geq n^{8}$): }

From the orthogonality of the Krawtchouk polynomials we have the bound
\begin{equation}
\big(K_{i}^{d(n-k)}(dl)\big)^2 \leq 2^{d(n-k)}\binom{d(n-k)}{i}\binom{d(n-k)}{dl}^{-1}.
\end{equation}
Inserting this into the expression above and using the triangle inequality we get
\begin{multline}
M_p(n,k,l,d) = \sum_{\substack{i=0\\i=dk \mod 2}}^{dk}| K^{d(n-k)}_i(dl) |\frac{(dk)! (d(n-k) - i-1)!!}{(dk-i)!! (dn-1)!!}\\
\le \sum_{\substack{i=0\\i = dk \mod 2}}^{dk} 2^{d(n-k)/2}\binom{d(n-k)}{i}^{1/2}\binom{d(n-k)}{l}^{-1/2} \frac{(dk)!}{(dk-i)!!}\frac{ (d(n-k) - i - 1)!!}{(dn-1)!!}.
\end{multline}

To obtain an upper bound, we first upper-bound the sum over $i$. To this end, we use the following facts about double factorials and binomials
(for $n$ even)
\begin{align}
(n-1)!! &= \frac{n!}{2^{n/2} (n/2)!}\quad \Rightarrow\quad 
\frac{\sqrt{n!}}{(n-1)!!} = 2^{n/2} \binom{n}{n/2}^{-1/2}\\
\frac { 2^{n H(p)}}{n+1} &\le \frac { 2^{n H(p)}}{\sqrt{8 p (1-p) n }} \le \binom{n}{pn}  \le  \frac { 2^{n H(p)}}{\sqrt{2\pi p (1-p) n }} \le 2^{n H(p)}, 
\label{eq:entropy bound binomial}
\end{align}
where $p \in [0,1]$ and $H(p) = - p \log( p) - (1-p) \log(1-p)$ is the binary entropy with base $2$
We then obtain
\begin{align}
    \binom{d(n-k)}{i}^{1/2} &\frac{(d(n-k)-i-1)!!}{(dk-i)!!} \notag\\&=\left(\frac{(d(n-k))!}{(d(n-k)-i)!i!} \right)^{1/2}\frac{(d(n-k)-i-1)!!}{(dk-i)!!}\\
    &\leq2^{-(d(n-k)-i)/2} \binom{d(n-k)-i}{(d(n-k)-i)/2}^{1/2} \sqrt{\frac{(d(n-k))!}{i!(dk-i)!}}\\
    &\le 2^{-(d(n-k)-i)/2} \frac{2^{(d(n-k)-i)/2}}
    {\sqrt{\pi (d(n-k)-i)}} \sqrt{\frac{(d(n-k))!}{i!(dk-i)!}}\\
    &\leq \sqrt{(d(n-k))!}  \frac{1}{\sqrt{i!(dk-i)!}},
\end{align}
where we used that $(dk-i)!! \ge \sqrt{(dk-i)!}$ and that $h(1/2) = 1$. 
Now we observe that this term is maximized at $i = \lfloor dk/2\rfloor $, since $i! (dk-i)!$ is minimized at $i=\lfloor dk/2\rfloor$. 
We thus get
\begin{equation}
\binom{d(n-k)}{i}^{1/2} \frac{(d(n-k)-i-1)!!}{(dk-i)!} \le \frac{((d(n-k))!)^{1/2} }{\lfloor dk/2\rfloor !}.
\end{equation}

Hence we can upper bound
\begin{multline}
M_p(n,k,l,d) = \sum_{\substack{i=0\\i=dk \mod 2}}^{dk} |K^{d(n-k)}_i(dl)| \frac{(dk)! (d(n-k) - i-1)!!}{(dk-i)!! (dn-1)!!}\\
\le 2^{d(n-k)/2}  \binom{d(n-k)}{dl}^{-1/2} dk \frac{(dk)!}{\lfloor dk/2\rfloor!}\frac{((d(n-k))!)^{1/2} }{(dn-1)!!}.
\end{multline}
We can insert this back into \Cref{eq:expression for average m2} and get
\begin{align}
 2^n|&T_{\mathrm{Case}\;2}| \notag\\
 &\le 6 \hspace{-.75cm}\sum_{\substack{(k,l)\in \mc S,\\8\log(n)\leq k\leq l\leq n-k-l}}\binom{n}{k}\binom{n-k}{l}\left[\sum_{\substack{i=0\\i=dk \mod 2}}^{dk} |K^{d(n-k)}_i(dl)| \frac{(dk)! (d(n-k) - i-1)!!}{(dk-i)!! (dn-1)!!}\right]\\
&\leq \hspace{-.75cm}\sum_{\substack{(k,l)\in \mc S,\\8\log(n) \le k\leq l\leq n-k-l}}\hspace{-1.8em}6 dk 2^{d(n-k)/2} \binom{n}{k}\binom{n-k}{l}\binom{d(n-k)}{dl}^{-\frac{1}{2}} \frac{(dk)!}{\lfloor dk/2 \rfloor !} \frac{((d(n-k))!)^{\frac{1}{2}} }{(dn-1)!!}
\\ 
&\le\hspace{-.75cm}\sum_{\substack{(k,l)\in \mc S,\\8\log(n) \le k\leq l\leq n-k-l}}\hspace{-1.8em}6 d^{3/2}k\sqrt{n} 2^{d(n-k)/2} \binom{n}{k}\binom{n-k}{l}\binom{d(n-k)}{dl}^{-\frac{1}{2}} \binom {dn}{dk}^{-\frac{1}{2}}  \\
&\le\hspace{-.75cm} \sum_{\substack{(k,l)\in \mc S,\\8\log(n) \le k\leq l\leq n-k-l}}\hspace{-1.8em}C d^{3/2}k\sqrt{n} 2^{dn/2} \binom{n}{k}\binom{n-k}{l}\binom{d(n-k)}{dl}^{-\frac{1}{2}} \binom {dn}{dk}^{-\frac{1}{2}}  \\
&\le\hspace{-.75cm} \sum_{\substack{(k,l)\in \mc S,\\8\log(n) \le k\leq l\leq n-k-l}}\hspace{-1.8em}C d^{3/2}k\sqrt{n} 2^{dn/2} \binom{n}{k}\binom{n-k}{l}\binom{(n-k)}{l}^{-\frac{d}{2}} \binom {n}{k}^{-\frac{d}{2}}
\label{eq:upper bound k geq 8 log n}
\end{align}
where we used $(dn-1)!!\geq \sqrt{(dn-1)!}  = (dn)^{-1/2} \sqrt{dn}$ and $C>6$ is a constant, and $\binom {dn}{dk} \geq \binom {n}{k}^d$ in the final line. By construction, and $d\geq 3$, we now have
\begin{align}
2^{-n(1-d/2) }  \binom{(n-k)}{l}^{1-\frac{d}{2}} \binom {n}{k}^{1-\frac{d}{2}} \leq n^{-8d/2 + 8} \leq n^{-4}.
\end{align}
This makes the total sum

\begin{equation}
|T_{\mathrm{Case}\;2}|\in O(n^{-1}) = o(1)
\end{equation}
\paragraph{Case 3 ($k\geq 90\log(n)$ and $n^{-3}\le 2^{-n} \binom{n}{k} \binom{n-k}{l} \leq n^{8}$): }
This case is more subtle and corresponds to the dominant terms of the sum over $k,l$ which lie in a `ring' of intermediate values, see \cref{fig:ring plot}(right). 
The first thing we note is that within this region there is a maximal value of $k$, achieved when $l = k$ and
\begin{equation}
2^{-n} \binom{n}{k^*} \binom{n-k^*}{k^*}  = n^8.
\end{equation}
We can (numerically) invert this equation to show that
\begin{equation}
k^* \leq cn + C_1 \log{n} \le c^* n,
\end{equation}
for some constant $C_1$ and $c_* \coloneqq 0.113<c <0.114 \eqqcolon c^*$ and sufficiently large $n$. Similarly we can show that there is a minimal $l_*$ given by
\begin{equation}
l_* \ge cn - C_2 \log{n} \ge c_*n,
\end{equation}
for some constant $C_2$ and sufficiently large $n$. 
The minimal $k_* = 90\log(n)$ further implies a maximal $l^* = (n-90\log(n))/2$.

Now we can use Derksen's bound:
\begin{multline} 
M_p(n,k,l,d) = \sum_{\substack{i=0\\i=dk \mod 2}}^{dk}  |K^{d(n-k)}_i(dl)| \frac{dk! (d(n-k) - i-1)!!}{(dk-i)!! (dn-1)!!}\\
\leq\sum_{\substack{i=0\\i = dk \mod 2}}^{dk} \binom{d(n-k)}{i}\bigg(\frac{i}{d(n-k)} + \frac{(n-k-2l)^2}{(n-k)^2}\bigg)^{i/2}\frac{(dk)!(d(n-k) - i - 1)!!}{(dk-i)!!(dn-1)!!}.
\label{eq:derksen matchings}
\end{multline}
This equation is manifestly monotonously decreasing in $l$, since $l\leq (n-k)/2$ for any pair $(k,l)$. 
Hence we can insert $l_*$ to obtain an upper bound on \cref{eq:derksen matchings}. 
We further split the sum over $i$ into two cases: (i) $i\geq 130\log(n)$ and (ii) $i\leq 130\log(n)$. 
For case (i) we can use H\"older's inequality 
\begin{align}
&\sum_{\substack{i\ge 130 \log n\\i = dk \mod 2}}^{dk} \binom{d(n-k)}{i}\bigg(\frac{i}{d(n-k)} + \frac{(n-k-2l)^2}{(n-k)^2}\bigg)^{i/2}\frac{(dk)!(d(n-k) - i - 1)!!}{(dk-i)!!(dn-1)!!}\\
&\leq \max_{\substack{ 130\log(n) \le i \le dk \\i = dk \mod 2}} \bigg(\frac{i}{d(n-k)} + \frac{(n-k-2l_*)^2}{(n-k)^2}\bigg)^{i/2} \!\!\!\!\sum_{\substack{i\geq 130\log(n)\\i = dk \!\!\mod 2}}^{dk}\!\!\!\!\! \binom{d(n-k)}{i}\frac{(dk)!(d(n-k) - i - 1)!!}{(dk-i)!!(dn-1)!!}\notag
\end{align}
We can upper bound the second factor by $1$ by noting that $\binom{d(n-k)}{i} = K_{i}^{d(n-k)}(0)$ and adding back in the $i\leq 130\log(n)$ terms to the sum. 
All of these terms are nonnegative since the only nonzero term at $x=0$ in the definition \eqref{eq:krawtchouk} of the Krawtchouk polynomials is the $q=0$ term.
This allows us to recognize the formula for the average parity over all matchings from \cref{lem:krawtchouk} with $l=0$ which trivially evaluates to $1$. For the maximization we note that $0 \le i/d\leq k\leq k^* = c^*n$ and $l_* \ge  c_*n$, telling us that
\begin{multline}
    \bigg(\frac{i}{d(n-k)} + \frac{(n-k-2l_*)^2}{(n-k)^2}\bigg)^{i/2} 
    \le \bigg(\frac{i}{d(n-k^*)} + \frac{(n-2l_*)^2}{(n-k^*)^2}\bigg)^{i/2} \\
    \leq \bigg( \frac {c^*}{1- c^*} + \frac{(1-2c_*)^2}{(1-c^*)^2}\bigg)^{i/2} 
    \le 0.892^{i/2} \le n^{-10.5},
\end{multline}
where the last bound follows from the fact that $0.892^{i/2}$ decreases in $i$, and hence we can use the lower bound $i \ge 130 \log n$ to obtain the upper bound. 

Finally we deal with the term where $i\leq 130 \log(n)$. This sub-case closely resembles Case~1 ($k \le 90 \log n$). 
We will use the straightforward bound
\begin{align}
\binom{d(n-k)}{i}&\leq (d(n-k))^i,
\end{align}
and furthermore bound 
\begin{align}
(d(n-k)-i-1)!! &= (d(n-k)-1)!! \prod_{t=0}^{i/2} (d(n-k)-2t-1)^{-1}\\
&\leq \frac{(d(n-k)-1)!!}{(d(n-k)-i-1)^{i/2}}\\
&\leq 
\frac{\sqrt{(d(n-k))!}}{(d(n-k))^{i/2}}(1+o(1)),
\end{align}
using that $i\le 130 \log n$ and therefore using Taylor's theorem $(d(n-k)-i-1)^{-i/2} \le (d(n-k))^{-i/2}( 1 +O(\log^2(n)/(dn))) \le (d(n-k))^{-i/2}( 1 +o(1)) $.  
Analogously, we find 
\begin{align}
  (dk-i)!!&\geq dk!! \prod_{t = 1}^{i/2} (dk-2t)^{-1} \geq (dk)!! (dk)^{-i/2}\geq \sqrt{(dk)!}(dk)^{-i/2},
\end{align}
Filling these in we obtain for case (ii)
\begin{align}
&\sum_{\substack{i=0\\i = dk \mod 2}}^{\lfloor 130\log(n)\rfloor} \binom{d(n-k)}{i}\bigg(\frac{i}{d(n-k)} + \frac{(n-k-2l_*)^2}{(n-k)^2}\bigg)^{i/2}\frac{(dk)!(d(n-k) - i - 1)!!}{(dk-i)!!(dn-1)!!}\\
&\le \frac{ \sqrt{(dk)!(d(n-k))! }}{\sqrt{(dn-1)!}} \sum_{\substack{i=0\\i = dk \mod 2}}^{\lfloor 130\log(n)\rfloor} \left[\frac{(dk)(d(n-k))^2}{d(n-k)}  \bigg(\frac{i}{d(n-k)} + \frac{(n-k-2l_*)^2}{(n-k)^2}\bigg) \right]^{i/2} (1+o(1))\notag\\
&\le \frac{(dn) \sqrt{(dk)!(d(n-k))! }}{\sqrt{(dn)!}} \sum_{\substack{i=0\\i = dk \mod 2}}^{\lfloor 130\log(n)\rfloor} \left[d^2k(n-k)  \bigg(\frac{i}{d(n-k)} + \frac{(n-k-2l_*)^2}{(n-k)^2}\bigg) \right]^{i/2} (1+o(1))]\notag
\end{align}
For the second factor in the sum we can argue that
\begin{equation}
\bigg(\frac{i}{d(n-k)} + \frac{(n-k-2l_*)^2}{(n-k)^2}\bigg)^{i/2} \leq \left[o(1) + \left(\frac{1-2c_*}{1-c^*}\right)^{2}\right]^{i/2} \le 1,
\end{equation}
for sufficiently large $n$, using the upper bounds on $i$ and $k$ and the lower bound on $l$. Furthermore the first factor is clearly bounded by 
\begin{equation}
 (d^2 k (n-k))^{i/2}\leq (dn)^i\leq (dn)^{130\log(n)}. 
\end{equation}
Combining all of these we get
\begin{multline}
\sum_{\substack{i=0\\i = dk \mod 2}}^{\lfloor 130\log(n)\rfloor} \binom{d(n-k)}{i}\bigg(\frac{i}{d(n-k)} + \frac{(n-k-2l_*)^2}{(n-k)^2}\bigg)^{i/2}\frac{(dk)!(d(n-k) - i - 1)!!}{(dk-i)!!(dn-1)!!}\\
\leq \binom{dn}{dk}^{-1/2} (dn)^{130\log(n) +1/2}(1+o(1)). 
\end{multline}
We can further bound this as 
\begin{align}
 &\hspace{-3cm}\log (\binom{dn}{dk}^{-1/2} (dn)^{130\log(n) +1/2}(1+o(1))) \\
&\le -dn H(k_*/n)/2 + 130 (\log^2(n) + \log n \log d ) + o(1)\\
&\le -d k_* \log (n/k_*)/2 + 130 (\log^2(n) + \log n \log d ) + o(1) \\
&\le \log^2 (n) (-45 d + 130 + o(1)) \\
&\le - 5 \log^2(n)(1 - o(1))
\end{align}
where we have used that $1/\binom{dn}{d k}$ is monotonously decreasing in $k$, so that we can bound the binomial using lower bound $k_* = 90\log(n)$, and the last bound follows from the assumption $2 < d = o(n^{1/2})$. Hence we have that
\begin{equation}
|T_{\mathrm{Case}\;3}|\le  n^2 n^{8} \cdot O(n^{-10.5} + n^{-\Omega(\log n)})  
\le  O(n^{-1/2}) = o(1).
\end{equation}
which completes the final case. 

\bigskip

Putting the three bounds together, we have our result that for $2 < d \in o(n^{1/2})$
\begin{align}
   \mb E_{G_M\sim \mc G_p(n,d),\theta} \left[m_2(G_M, \theta)\right]  = o(1) + \begin{cases}
       2& \text{ if } d = 1 \mod 2 \\ 
       3 & \text{ if } d = 0 \mod 2.
   \end{cases}
\end{align}
\end{proof}
\begin{proof}[Sketch of proof for the random matching case of  \Cref{thm:anticoncentration}]
\label{proof:anticoncentration matching}
Let us conclude this section by sketching the proof of the random matchings part of \Cref{thm:anticoncentration}. 
The proof proceeds along the same lines as \Cref{thm:anticoncentration random pairing} but is somewhat simpler. We restrict ourselves to outlining the essential difference and leaving the details of adapting the proof as an exercise to the reader. \\

The key difference in the proof is the starting point, where instead of averaging over one large matching of $dn$ vertices, we average over $d$ independent matchings of $n$ vertices. 
\begin{align}
\hspace{-1ex}\mb E_{G\sim \mc G_m(n,d),\theta}(m_2(G, \theta) &= 2^{-n}\sum_{L,R\subset [n], L\cap R = \emptyset} \mathbb{E}_{G\sim \mc G_m(n,d)}(-1)^{|A_{G}[L,R]|} \\
&= 2^{-n}\sum_{L,R\subset [n], L\cap R  = \emptyset} \mathbb{E}_{M_1, \ldots, M_d \sim \mc M(n)}(-1)^{|A_{M_1}[L,R]|} \cdots (-1)^{|A_{M_d}[L,R]|}\notag\\
&= 2^{-n}\sum_{L,R\subset [n], L\cap R  = \emptyset} \left(\mathbb{E}_{M\sim \mc M(n)}(-1)^{|A_{M}[L,R]|} \right)^d,
\end{align}
and we thus get by \Cref{lem:symmetry,lem:krawtchouk}
\begin{multline}
\label{eq:anticoncentration random matching}
   \mb E_{G\sim \mc G_m(n,d),\theta}(m_2(G, \theta) = 2^{-n}\!\!\!\!\!\!\!\!\!\!\!\sum_{L,R\subset [n], L\cap R  = \emptyset} \left(\sum_{\substack{i=0\\i=|L| \mod 2}}^{|L|}\hspace{-.8em} K^{n-|L|}_i(|R|) \frac{|L|! (n-|L| - i-1)!!}{(|L|-i)!! (n-1)!!}\right)^d.\hspace{-.8em}
\end{multline}
In the subsequent analysis many aspects of the proof simplify since the $d$-dependence is isolated to an overall power of the sum over $i$. 
\end{proof}

\section{Universality of regular graphs of intermediate degree}\label{sec:intermediate}
In this section, we prove that most regular graphs of sufficiently high regularity are resources for universal quantum computation. We prove this by showing that we can find sufficiently large grid graphs embedded in these graphs as induced subgraphs. 

Our proof consists of a series of lemmas. We start with an exposition of the ``switching method'' of regular graph theory ---a counting technique first introduced by Mckay and Wormald \cite{mckay1990uniform}--- as it is applicable in our context. We use the switching method to compute the expected number of grid graphs in a random regular graph. After that, we compute the associated variance. Finally, we apply Chebyshev's inequality to make a typicality statement.  Our proofs follow the arguments for the estimation of the probability of subgraphs \cite{kim_small_2007}, and induced subgraphs \cite{xiao_induced_2008-1} \emph{of constant size} of random regular graphs. However we require estimates for induced subgraphs that grow in size with $n$, which means we need to be substantially more careful in our estimations.

\subsection{An exposition of forward and reverse switching}
\label{subsection: switching}
In our proofs in \Cref{sec:intermediate}, we make heavy use of a counting technique called switching, which was introduced by Mckay and Wormald \cite{mckay1990uniform}. This is a standard technique in the regular-graph theory literature (and combinatorics more generally), see e.g. \cite{hasheminezhad2010combinatorial} for a general discussion. Because we need somewhat precise estimates, and this technique is not very well known in the quantum information literature, we do several key calculations explicitly here (specifically for our problem).\\

Let $\varepsilon, \varepsilon'$ be two collections of edges on the vertex set $[n]$, with $\varepsilon \cap \varepsilon' = \emptyset$ and $|\varepsilon| = s,|\varepsilon'| = s' $. Let $uw$ be an edge in $\varepsilon$, and $u'w'$ be an edge in $\varepsilon'$. Define two sets:
\begin{align}
    \mathscr{L} &= \left\{ G_{n,d} \mid \varepsilon \subseteq G_{n,d}; \varepsilon' \subseteq \overline{G}_{n,d} \right\}, \\
     \mathscr{M} &= \left\{ G_{n,d} \mid \varepsilon \backslash \{uw\} \subseteq G_{n,d}, uw \notin G_{n,d}; \varepsilon' \subseteq \overline{G}_{n,d} \right\}.
\end{align}

The goal of the switching method is to estimate $|\mathscr{L}|/ |\mathscr{M}|$. This is done by setting up a relation $R$ between $\mathscr{L} ,\mathscr{M}$, and then estimating how many elements of $\mathscr{M}$ are related to a uniformly random element of $\mathscr{L}$ (call this expectation $d_1$). We will also estimate how many elements of $\mathscr{L}$ are related to a uniformly random element of $\mathscr{M}$ (call this $d_2$). By a basic double counting argument one can see that
\begin{equation}
d_1 |\mathscr{L}| = |R| = d_2 |\mathscr{M}|.
\end{equation}
This means we can estimate the ratio by estimating $d_1,d_2$ (note that this part is not specific to the sets in question). We now set up the relation (which is specific to these sets). We do this by relating a graph $G'\in \mathscr{M}$ to a graph $G \in \mathscr{L}$ if $G'$ can be reached from $G$ by a \emph{forward switching}. Equivalently, we will see that this means $G'$ can be reached from $G$ by \emph{reverse switching}.

Given a graph $G \in  \mathscr{L}$, we will choose two edges $u_1 w_1$ and $u_2 w_2$ of $G \backslash \varepsilon$, delete these edges together with the edge $uw$ and insert the new edges $wu_1, w_1 u_2, w_2 u$. This produces a graph $G'$. We will choose $u_1 w_1$ and $u_2 w_2$ of $G \backslash \varepsilon$ such that all six endpoints of the edges are distinct and, in addition, $wu_1, w_1 u_2, w_2 u$ are not edges of $G$ and not in $\varepsilon'$.

We can provide an estimate for how many ways this forward switching can be performed (and thus estimate $d_1$).

\begin{lemma}\label{lem:forward switching}
Given $\varepsilon, \varepsilon'$ defined as above, there are 
\begin{equation}
d^2(n - \mathcal{O}(d))^2\cdot\bigg(1 - \frac{\mathcal{O}(s)}{(n-\mathcal{O}(d)) \cdot d} - \frac{\mathcal{O}(s')}{n-\mathcal{O}(d)}  \bigg)
\end{equation}
ways to perform a forward switching from $\mathscr{L}$ to $\mathscr{M}$.
\end{lemma}
\begin{proof}
The number of choices of $u_1 w_1$ is 
\begin{equation}
2 \left(\frac{nd}{2} - \# \textsf{forbidden cases}\right),
\end{equation}
where $nd/2$ is the total number of edges of the graph and we multiply by $2$ because $u_1w_1$ and $w_1u_1$ represent two separate cases when inserting the new edges. The forbidden cases are summarized as follows:
\begin{itemize}
\item There are at most $s$ choices of $u_1w_1$ such that $u_1w_1$ is in $\epsilon$.
\item There are at most $(d - 1)$ choices such that $w = u_1$; that is, the endpoints will not be distinct. 
\item There are at most
\begin{equation}
\mathcal{O}\left((d-1)^2 + s'd\right)
\end{equation}
choices of $u_1w_1$ such that $wu_1$ and $w_1u_2$ are either in $G$ or in $\epsilon'$.
\end{itemize}
\noindent Hence,
\begin{equation}
\#\textsf{forbidden cases} = \mathcal{O}\bigg(s + (d-1) + (d-1)^2 + s'd\bigg) = \mathcal{O}\bigg( s + d^2 + s' d \bigg).
\end{equation}
\noindent Putting everything together, the number of choices of $u_1 w_1$ are
\begin{equation}
\label{eq: choices}
nd - \mathcal{O}\bigg(s + d^2 + s'd \bigg).
\end{equation}
By a similar argument, the number of choices of $u_2 w_2$ is also given by \Cref{eq: choices}. Hence, the total number of choices of $u_1w_1$ and $u_2w_2$ are 
\begin{equation}
\begin{aligned}
\label{eqn: forward switching}
&\left(d(n - \mathcal{O}(d)) - \mathcal{O}(s + s'd)\right)^2 \\
&=d^2(n - \mathcal{O}(d))^2\cdot\bigg(1 - \frac{\mathcal{O}(s + s'd)}{(n-\mathcal{O}(d)) \cdot d} + \frac{\mathcal{O}(s+s'd)^2)}{(n-\mathcal{O}(d))^2 \cdot d^2} \bigg) \\
&= d^2(n - \mathcal{O}(d))^2\cdot\bigg(1 - \frac{\mathcal{O}(s)}{(n-\mathcal{O}(d)) \cdot d} - \frac{\mathcal{O}(s')}{n-\mathcal{O}(d)}  \bigg).
\end{aligned}
\end{equation}
\end{proof}

We still need to estimate $d_2$. To do this we define an inverse operation called \emph{reverse switching}, mapping graphs in $\mathscr{M}$ to graphs in $\mathscr{L}$.

Starting from a graph $G' \in \mathscr{M}$ we delete edges $wu_1, w_1 u_2, w_2 u$ of $G' \backslash \varepsilon$ and insert edges $uw, u_1 w_1, u_2 w_2$. Again, we allow only switchings for which all six vertices are distinct, and $u_1 w_1, u_2 w_2$ are not edges of $G'$ and not in $\varepsilon'$. This operation produces a graph which belongs to $\mathscr{L}$. Note also that the existence of a reverse switching from $G'$ to $G$ implies the existence of a forward switching from $G$ to $G'$. Hence counting the number of reverse switchings for a random $G' \in \mathscr{M}$ is equivalent to calculating $d_2$.

We can again provide an estimate of the number of ways in which this reverse switching can be performed. 

\begin{lemma}\label{lem:reverse switching}
Given $\varepsilon, \varepsilon'$ defined as above, there are 
\begin{equation}
d^3 (n - \mathcal{O}(d)) 
\left(1 - \mathcal{O}\left(\frac{s}{d}\right) 
- \frac{\mathcal{O}(s')}{n-\mathcal{O}(d)} \right)
\end{equation}
ways to perform a reverse switching from $\mathscr{L}_3$ to $\mathscr{L}_1$.
\end{lemma}

\begin{proof}
Note that $u$ and $w$ are vertices in $G'$. Hence the number of choices of $wu_1$ and $w_2u$ such that they are in $G' \backslash \varepsilon$ are 
\begin{align}
\left(d - 1  - \mathcal{O}(s)\right)^2 = \left(d  - \mathcal{O}(s)\right)^2
\end{align}
The number of choices of $w_1 u_2$ is 
\begin{equation}
2 \left(\frac{nd}{2} - \# \textsf{forbidden cases}\right).
\end{equation}
The forbidden cases are summarized as follows:
\begin{itemize}
\item There are at most $s$ choices of $w_1u_2$ such that $u_1w_1$ is in $\epsilon$.
\item Having already picked $w, u_1, w_2$ and $u$, there are $\mathcal{O}(d - 1)$ choices such that the endpoints of $w_1u_2$ will not be distinct. 
\item There are at most
\begin{equation}
\mathcal{O}((d-1)^2 + s'd)
\end{equation}
choices of $w_1u_2$ such that $u_1w_1$ and $u_2w_2$ are either in $G'$ or in $\epsilon'$.
\end{itemize}
\noindent Hence,
\begin{equation}
\#\textsf{forbidden cases} = \mathcal{O}\bigg(s + (d-1) + (d-1)^2 + s'd\bigg) = \mathcal{O}\bigg( s + d^2 + s' d \bigg).
\end{equation}
\noindent Taken together, the total number of choices of $wu_1, w_1u_2$ and $w_2u$ are
\begin{equation}
\begin{aligned}
\label{eqn: reverse switching}
&\left(d - \mathcal{O}(s) \right)^2 
\left(d(n - \mathcal{O}(d)) - \mathcal{O}(s + s'd) \right) \\
&= \left(d - \mathcal{O}(s) \right)^2 
\left(d(n - \mathcal{O}(d)) - \mathcal{O}(s + s'd) \right) \\
&= d^3 (n - \mathcal{O}(d)) 
\left(1 - \frac{\mathcal{O}(s - 2ss')}{d \cdot (n-\mathcal{O}(d))} 
- \frac{\mathcal{O}(s')}{n-\mathcal{O}(d)} 
- \mathcal{O}\left(\frac{s}{d}\right) \right. \\
&\quad \left.
- \frac{\mathcal{O}(s's^2 - s^2)}{d^2 \cdot (n-\mathcal{O}(d))} 
+ \mathcal{O}\left(\frac{s^2}{d^2}\right) 
- \frac{\mathcal{O}(s^3)}{d^3 \cdot (n-\mathcal{O}(d))} \right) \\
&= d^3 (n - \mathcal{O}(d)) 
\left(1 - \mathcal{O}\left(\frac{s}{d}\right) 
- \frac{\mathcal{O}(s')}{n-\mathcal{O}(d)} \right).
\end{aligned}
\end{equation}
\end{proof}
\noindent From the description of the processes, the following corollary is immediate.
\begin{corollary}
    $G$ is reachable from $G'$ by forward switching if and only if $G'$ is reachable from $G$ by reverse switching.
\end{corollary}
\subsubsection{More examples}
\label{subsubsection: more examples}
\noindent We can also estimate the relative sizes of two other sets given by 
\begin{align}
\mathscr{L} &= \left\{ G_{n,d} \mid \varepsilon \backslash \{uw\} \subseteq G_{n,d}; \varepsilon' \backslash \{u'w'\} \subseteq \overline{G}_{n,d}, u'w' \in {G}_{n,d} \right\}, \\
\mathscr{M} &=  \left\{ G_{n,d} \mid \varepsilon \backslash \{uw\} \subseteq G_{n,d}; \varepsilon' \subseteq \overline{G}_{n,d} \right\}
\end{align}
using forward and reverse switching. Just as before, given a graph $G \in  \mathscr{L}$, we will choose two edges $u_1 w_1$ and $u_2 w_2$ of ${G} \backslash \varepsilon$ such that all six endpoints $u', w', u_1, w_1, u_2,$ and $w_2$ are distinct and, in addition, $w'u_1, w_1 u_2, w_2 u'$ are not edges of ${G}$ and not in $\varepsilon'$. Then, we will forward switch to get a graph $G'$ in $\mathscr{M}$. Similarly, we can reverse switch to go from a graph in $\mathscr{M}$ to a graph in $\mathscr{L}$.

\subsection{Expected number of induced subgraphs}
We plan to use the second moment method to prove \Cref{thm:grid_subgraph}. In order to do this we first calculate the expected number of graphs isomorphic to a fixed graph $H$ of a given size that show up as induced subgraphs of a randomly sampled $d$ regular graph. In the next subsection we will then specialize to grid graphs. Throughout this graph will be represented by a set of edges $\varepsilon$ and a set of non-edges $\varepsilon'$. This latter part is required because we want to find \emph{induced} subgraphs.

In order to compute the expected number of induced subgraphs we first need to prove a few subsidiary lemmas. The first allows us to ``peel off'' edges from the sets $\varepsilon,\varepsilon'$, without changing the probability of observing these sets too much. This lemma explicitly relies on the switching technique:

\begin{lemma}
\label{lemma: regular graphs of intd. degree}
Let $d = n^{c}$, for any choice of constant $0 < c < 1$, and let $\varepsilon, \varepsilon'$ be two collections of edges on the vertex set $[n]$, with $\varepsilon \cap \varepsilon' = \emptyset$ and $|\varepsilon| = s,|\varepsilon'| = s'$. Let $uw$ be an edge in $\varepsilon$, and $u'w'$ be an edge in $\varepsilon'$. Then, we have:
\begin{multline}
    \mathbb{P}\left[\varepsilon \subseteq G_{n,d}, \varepsilon' \subseteq \overline{G}_{n,d}\right] \\ = \left(\frac{n\alpha}{n - {d(1 - \alpha)}}\right) \left( \frac{d}{n} \right) \left( 1 - \frac{d}{n} \right) \mathbb{P}\left[\varepsilon \backslash \{uw\} \subseteq G_{n,d}, \varepsilon' \backslash \{u'w'\} \subseteq \overline{G}_{n,d}\right],
\end{multline}
\noindent where
\begin{equation}
\alpha = \frac{1 - \mathcal{O}\left(\frac{s}{d}\right) - \frac{\mathcal{O}(s')}{n-\mathcal{O}(d)}}{1 - \frac{\mathcal{O}(s)}{(n-\mathcal{O}(d)) \cdot d} - \frac{\mathcal{O}(s')}{n-\mathcal{O}(d)}}.
\end{equation}
\end{lemma}

\begin{proof}
Note that $uw \in \epsilon$ and $u'w' \in \epsilon'$. Let
\begin{align}
    \mathscr{L}_1 &= \left\{ G_{n,d} \mid \varepsilon \backslash \{uw\} \subseteq G_{n,d}; \varepsilon' \backslash \{u'w'\} \subseteq \overline{G}_{n,d}; uw \in G_{n,d}; u'w' \in \overline{G}_{n,d} \right\}, \\
    \mathscr{L}_2 &= \left\{ G_{n,d} \mid \varepsilon \backslash \{uw\} \subseteq G_{n,d}; \varepsilon' \backslash \{u'w'\} \subseteq \overline{G}_{n,d}; uw \in G_{n,d}; u'w' \notin \overline{G}_{n,d} \right\}, \\
    \mathscr{L}_3 &= \left\{ G_{n,d} \mid \varepsilon \backslash \{uw\} \subseteq G_{n,d}; \varepsilon' \backslash \{u'w'\} \subseteq \overline{G}_{n,d}; uw \notin G_{n,d}; u'w' \in \overline{G}_{n,d} \right\}, \\
    \mathscr{L}_4 &= \left\{ G_{n,d} \mid \varepsilon \backslash \{uw\} \subseteq G_{n,d}; \varepsilon' \backslash \{u'w'\} \subseteq \overline{G}_{n,d}; uw \notin G_{n,d}; u'w' \notin \overline{G}_{n,d} \right\}.
\end{align}

\noindent By suppressing the redundancies in notation, we can rewrite the same equations more succinctly as
\begin{align}
    \mathscr{L}_1 &= \left\{ G_{n,d} \mid \varepsilon \subseteq G_{n,d}; \varepsilon' \subseteq \overline{G}_{n,d} \right\}, \\
    \mathscr{L}_2 &= \left\{ G_{n,d} \mid \varepsilon \subseteq G_{n,d}; \varepsilon' \backslash \{u'w'\} \subseteq \overline{G}_{n,d}, u'w' \in G_{n,d} \right\}, \\
    \mathscr{L}_3 &= \left\{ G_{n,d} \mid \varepsilon \backslash \{uw\} \subseteq G_{n,d}, uw \notin G_{n,d}; \varepsilon' \subseteq \overline{G}_{n,d} \right\}, \\
    \mathscr{L}_4 &= \left\{ G_{n,d} \mid \varepsilon \backslash \{uw\} \subseteq G_{n,d}, uw \notin G_{n,d}; \varepsilon' \backslash \{u'w'\} \subseteq \overline{G}_{n,d}, u'w' \in G_{n,d} \right\}.
\end{align}
By definition, we have that
\begin{align}    
\frac{\operatorname{Pr}\left[\varepsilon \subseteq G_{n,d}, \varepsilon' \nsubseteq G_{n,d}\right]}{\operatorname{Pr}\left[\varepsilon \backslash \{uw\} \subseteq G_{n,d}, \varepsilon' \backslash \{u'w'\} \subseteq \overline{G}_{n,d}\right]} &= \frac{\left| \mathscr{L}_1 \right|}{\left| \mathscr{L}_1 \right| + \left| \mathscr{L}_2 \right| + \left| \mathscr{L}_3 \right| + \left| \mathscr{L}_4 \right|} \\&= \frac{\frac{\left| \mathscr{L}_1 \right|}{|\mathscr{L}_1| + |\mathscr{L}_3|}}{1 + \frac{\left| \mathscr{L}_2 \right| + \left| \mathscr{L}_4 \right|}{\left| \mathscr{L}_1 \right| + \left| \mathscr{L}_3 \right|}}.
\end{align}

\noindent To prove the lemma, it is enough to show that
\begin{equation}
\label{thiseq}
    \frac{\left| \mathscr{L}_1 \right|}{\left| \mathscr{L}_3 \right|} =  \frac{\alpha \cdot d}{n - d} \quad \text{and} \quad \frac{\left| \mathscr{L}_2 \right| + \left| \mathscr{L}_4 \right|}{\left| \mathscr{L}_1 \right| + \left| \mathscr{L}_3 \right|} =\frac{ \alpha \cdot d}{n - d}.
\end{equation}

\noindent We first prove that $\frac{\left| \mathscr{L}_1 \right|}{\left| \mathscr{L}_3 \right|} = \frac{\alpha d}{n - d}$. Given a graph $G \in \mathscr{L}_1$, we will perform forward switching to get a graph in $\mathscr{L}_3$. By Choose two edges $u_1 w_1$ and $u_2 w_2$ of $G \backslash \varepsilon$, delete these edges together with the edge $uw$ and insert the new edges $wu_1, w_1 u_2, w_2 u$. We will choose $u_1 w_1$ and $u_2 w_2$ of $G \backslash \varepsilon$ such that all six endpoints of the edges are distinct and, in addition, $wu_1, w_1 u_2, w_2 u$ are not edges of $G$ and not in $\varepsilon'$. Then, it is easy to see that the graph obtained from $G$ by a forward switching belongs to $\mathscr{L}_3$. By \Cref{eqn: forward switching}, the number of possible forward switchings is
\begin{equation}
d(n - \mathcal{O}(d))\cdot\bigg(1 - \frac{\mathcal{O}(s)}{(n-\mathcal{O}(d)) \cdot d} - \frac{\mathcal{O}(s')}{n-\mathcal{O}(d)}  \bigg).
\end{equation}
Similarly, by applying reverse switching, we can go from $\mathscr{L}_3$ to $\mathscr{L}_1$. By \Cref{eqn: reverse switching}, the number of possible reverse switchings is
\begin{equation}
d^3 (n - \mathcal{O}(d)) \left(1 - \mathcal{O}\left(\frac{s}{d}\right) - \frac{\mathcal{O}(s')}{n-\mathcal{O}(d)} \right).
\end{equation}

\noindent Since $\mathscr{L}_2 \cap \mathscr{L}_4 = \emptyset$ and $\mathscr{L}_1 \cap \mathscr{L}_3 = \emptyset$, the ratio of $\left| \mathscr{L}_2 \right| + \left| \mathscr{L}_4 \right|$ and $\left| \mathscr{L}_1 \right| + \left| \mathscr{L}_3 \right|$ is
\begin{equation}
    \frac{\left| \mathscr{L}_2 \right| + \left| \mathscr{L}_4 \right|}{\left| \mathscr{L}_1 \right| + \left| \mathscr{L}_3 \right|} = \frac{\left| \mathscr{L}_2 \cup \mathscr{L}_4 \right|}{\left| \mathscr{L}_1 \cup \mathscr{L}_3 \right|}
\end{equation}

\noindent Moreover,
\begin{align}
    \mathscr{L}_2 \cup \mathscr{L}_4 &= \left\{ G_{n,d} \mid \varepsilon \backslash \{uw\} \subseteq G_{n,d}; \varepsilon' \backslash \{u'w'\} \subseteq \overline{G}_{n,d}, u'w' \notin \overline{G}_{n,d} \right\}, \\
    \mathscr{L}_1 \cup \mathscr{L}_3 &= \left\{ G_{n,d} \mid \varepsilon \backslash \{uw\} \subseteq G_{n,d}; \varepsilon' \subseteq \overline{G}_{n,d} \right\}.
\end{align}

\noindent
Once again, by the forward and reverse switching arguments as in \Cref{subsubsection: more examples}, we can show that
\begin{equation}
    \frac{\left| \mathscr{L}_2 \cup \mathscr{L}_4 \right|}{\left| \mathscr{L}_1 \cup \mathscr{L}_3 \right|} = \frac{\alpha d}{n - d}.
\end{equation}
\noindent Hence, the lemma follows.
\end{proof}
Next we need a useful intermediate result regarding the probability of finding a subgraph (not induced) in a random regular graph. The proof is a modification to that in \cite{kim_small_2007}. We will use this result in the proof of our main result.
\begin{lemma}
\label{lemma: probability_subgraph}
Let $d = n^c$ for any choice of $0 < c < 1$. Let $\varepsilon$ be a fixed disjoint collection of edges on the vertex set $[n]$ of size $s$ and $uw$ be an edge in $\epsilon$. Then,
\begin{equation}
\Pr[\varepsilon \subseteq G_{n,d}] = \left(\frac{n \alpha}{n - d(1-\alpha)}\right) \left(\frac{d}{n}\right) \Pr[\varepsilon \backslash \{uw\} \subseteq G_{n,d}],
\end{equation}
\noindent where
\begin{equation}
\alpha = \frac{1 - \mathcal{O}\left(\frac{s}{d}\right)}{1 - \frac{\mathcal{O}(s)}{(n-\mathcal{O}(d)) \cdot d}}.
\end{equation}
\end{lemma}

\begin{proof}
Let us define two sets
\begin{equation}
\mathscr{L}_0 = \{\varepsilon \subseteq G_{n,d} \},
\end{equation}
\begin{equation}
\mathscr{L}_1 = \{\varepsilon \backslash \{uw\} \subseteq G_{n,d} : \{uw\} \notin G_{n,d} \}.
\end{equation}
Given $G \in \mathscr{L}_0$, we apply a forward switching operation to get a graph in $\mathscr{L}_1$, and given a graph $G \in \mathscr{L}_1$, we apply reverse switching to get a graph in $\mathscr{L}_0$. By similar arguments as in \Cref{subsection: switching},
\begin{equation}
\frac{|\mathscr{L}_0|}{|\mathscr{L}_1|} = \frac{\alpha \cdot d}{n-d}.
\end{equation}
\noindent Now,
\begin{equation}
\label{subgraph_eq}
\frac{\Pr[\varepsilon \subseteq G_{n,d}]}{\Pr[\varepsilon \backslash \{uw\} \subseteq G_{n,d}]} = \frac{|\mathscr{L}_0|}{|\mathscr{L}_0| + |\mathscr{L}_1|} = \frac{\frac{|\mathscr{L}_0|}{|\mathscr{L}_1|}}{1+ \frac{|\mathscr{L}_0|}{|\mathscr{L}_1|}} = \frac{d}{n}  \left(\frac{n \alpha}{n - d(1-\alpha)}\right).
\end{equation}
\noindent From \Cref{subgraph_eq}, the lemma follows.
\end{proof}
\begin{corollary}
\label{corollary: subgraph}
Let $d = n^c$ for any choice of $0 < c < 1$. Let $uw$ be an edge. Then,
\begin{equation}
\Pr[uw \notin G_{n,d}] = \left(1 + \frac{o(d)}{n}\right) \left(1 - \frac{d}{n}\right).
\end{equation}
\end{corollary}
\begin{proof}
Follows from \Cref{lemma: probability_subgraph} by setting $s = 1$.
\end{proof}
With \Cref{lemma: probability_subgraph} and \Cref{lemma: regular graphs of intd. degree} we can estimate the probability that we observe an induced subgraph on a fixed set of vertices. 

\begin{lemma}
\label{lemma: probability_induced_subgraph}
Let $d = n^c$ for any choice of $0 < c < 1$. Let $\varepsilon, \varepsilon'$ be fixed disjoint collections of edges on the vertex set $[n]$, of size $s, s'$ respectively. If we assume that $s \cdot s' = o(d)$, then:
\begin{equation}
\Pr[\varepsilon \subseteq G_{n,d}, \varepsilon' \subseteq \overline{G}_{n,d}] = \left( 1 - o(1) \right) \left( \frac{d}{n} \right)^s \left( 1 - \frac{d}{n} \right)^{s'}.
\end{equation}
\end{lemma}
\begin{proof}
Without loss of generality, let $s < s'$. By recursively applying \Cref{lemma: regular graphs of intd. degree}, we have that
\begin{equation}
\begin{aligned}
&\Pr[\varepsilon \subseteq G_{n,d}, \varepsilon' \subseteq \overline{G}_{n,d}] \\&= \left(\frac{n\alpha}{n - {d(1 - \alpha)}}\right)^s \left( \frac{d}{n} \right)^s \left( 1 - \frac{d}{n} \right)^{s} \mathbb{P}\left[\varepsilon' \backslash \{u_1'w_1', u_2'w_2', \ldots, u_s'w_s'\} \subseteq \overline{G}_{n,d}\right],
\end{aligned}
\end{equation}
for choices of edges 
\begin{equation}
\{u_1'w_1', u_2'w_2', \ldots, u_s'w_s'\} \in \epsilon'
\end{equation}
that are picked at each round, where
\begin{equation}
\alpha = \frac{1 - \mathcal{O}\left(\frac{s}{d}\right) - \frac{\mathcal{O}(s')}{n-\mathcal{O}(d)}}{1 - \frac{\mathcal{O}(s)}{(n-\mathcal{O}(d)) \cdot d} - \frac{\mathcal{O}(s')}{n-\mathcal{O}(d)}}.
\end{equation}
\noindent When $s \cdot s' = o(d)$,
\begin{equation}
\left(\frac{n\alpha}{n - {d(1 - \alpha)}}\right)^s  = (1 - o(1)).
\end{equation}
\noindent Let the remaining edges in $\epsilon'$ be
\begin{equation}
\{ u'_{s+1}w'_{s+1}, \ldots u'_{s'}w'_{s'} \}.
\end{equation}
\noindent So, by \Cref{corollary: subgraph},
\begin{equation}
\label{eqn: to justify}
\mathbb{P}\left[\{ u'_{s+1}w'_{s+1}, \ldots u'_{s'}w'_{s'} \} \in \overline{G}_{n,d}\right] = \left(1 + \frac{o(d \cdot (s'-s))}{n} \right) \left(1 - \frac{d}{n}\right)^{s' - s}.
\end{equation}
When $ss' = o(d)$,
\begin{equation}
\left(1 + \frac{o(d \cdot (s'-s))}{n} \right) = 1 + o(1).
\end{equation}
From this observation, the lemma follows.
\end{proof}
 With these lemmas, we can finally prove the main result of this subsection, namely an estimate of the expected number of induced subgraphs isomorphic to a fixed graph $H$ (the size of which possible grows with $n$).
\begin{corollary}
\label{corollary: induced subgraph}
Let $d = n^c$ for any choice of $0 < c < 1$. Let $H$ be a fixed graph with $e$ edges and $v$ vertices with $e\left(\binom{v}{2} - e\right) = o(d)$; $Y_H$ denote the number of induced copies of $H$ in $G_{n,d}$ and $\textsf{aut}(H)$ be the number of automorphisms of $H$. Then, 
\begin{equation}
\label{expected_value}
\begin{aligned}
\mathbb{E}(Y_H) &= \left(1 - o(1)\right) \frac{\binom{n}{v} v!}{\textsf{aut}(H)} \left(\frac{d}{n}\right)^e \left(1 - \frac{d}{n}\right)^{\binom{v}{2} - e} 
\\&= \Theta\left( n^{v - e} d^e \left(1 - \frac{d}{n}\right)^{\binom{v}{2} - e} \right).
\end{aligned}
\end{equation}
\end{corollary}

\begin{proof}
For each copy $H'$ of $H$, define the indicator random variable $J_{H'}$ such that $J_{H'} = 1$ if and only if $H'$ is an induced subgraph of $G_{n,d}$. By Lemma \ref{lemma: probability_induced_subgraph}, 
\begin{equation}
\mathbb{E}(J_{H'}) = \left(1 - o(1)\right) \left(\frac{d}{n}\right)^e \left(1 - \frac{d}{n}\right)^{\binom{v}{2} - e}.
\end{equation}
Moreover, there are exactly $\frac{\binom{n}{v} v!}{\text{aut}(H)}$ copies of $H$, and by the linearity of expectation, \Cref{expected_value} holds.
\end{proof}

\subsection{Expected number of grid graphs}

Here we will specialize the results of the previous subsection to grid graphs. For a square grid on $v$ vertices (where we assume $v$ to be a square), the number of edges
\begin{equation}
\label{eq: edges of a grid}
e = 2v - 2\sqrt{v}.
\end{equation}
The number of edges in the complement of the grid graph is
\begin{equation}
\label{eq: edges of complement of grid}
\overline{e} = \frac{v^2 - 5v}{2} + 2\sqrt{v}.
\end{equation}
\noindent Hence, the condition
\begin{equation}
\left(\binom{v}{2} - e\right) e = o(d)
\end{equation}
of \Cref{corollary: induced subgraph} implies $v = o(d^{1/3})$.
\begin{corollary}\label{cor:expected_grid}
Let $d = n^c$ for any choice of $0 < c < 1$. Let $G_{n,d}$ be a random $d$-regular graph on $n$ vertices and let $H$ be a grid graph having $v$ vertices for any $v$ satisfying $v^2 \cdot d = \mathcal{O}(n)$ and $v = o(d^{1/3})$. Let $Y_H$ denote the number of induced copies of $H$ in $G_{n,d}$. Then, 
\begin{equation}
\label{eq: exp value}
\mathbb{E}(Y_H) = \Theta\left(\left(\frac{d^2}{n}\right)^v \right).
\end{equation}
\end{corollary}
\begin{proof}
By applying \Cref{corollary: induced subgraph}, 
\begin{equation}
\begin{aligned}
\mathbb{E}(Y_H) &= \Theta\left( n^{v - e} d^e \left(1 - \frac{d}{n}\right)^{\binom{v}{2} - e} \right). 
\end{aligned}
\end{equation}

\noindent The proof follows from combining together the following observations: $(1):v - e = -v + 2\sqrt{v} = -\Theta(v).$, $(2):n^{v-e} = n^{-\Theta(v)}$, $(3): d^{e} = d^{\Theta(2v)},$ and $(4):\left( 1 - \frac{d}{n}\right)^{\binom{v}{2} - e} = e^{-\Theta(v^2 d/n)}.$
\end{proof}

\noindent Observe that \Cref{eq: exp value} is greater than $1$ when $d = \omega(\sqrt{n})$.

\subsection{Expected number of sparsified square grid graphs}
\label{subsection: sparsified grid}
In the previous section we noted that grid graphs are expected to appear whenever $d  = \omega(\sqrt{n})$. We can push this lower bound on the degree further down by considering instead \emph{sparsified} grid graphs. These are constructed from regular $L\times L $ grid graphs by replacing each edge with $L-1$ vertices and $L$ edges connected in a line. These graphs, on $2L(L-1)^2 + L^2$ vertices, are still universal resource states, as measuring all but one of the qubits on each line in the $Y$ basis, which is equivalent to applying the local complementation operation on that vertex, gives back the $L\times L$ grid graph.  Choosing the number of vertices $v =2L(L-1)^2 + L^2$ , the number of edges in a sparsified grid graph is $e =2L^2(L-1)$. The number of edges in the complement is 
\begin{equation}
\overline{e} = \binom{2L(L-1)^2 + L^2}{2} - 2L^2(L-1).
\end{equation}
With this information we can reprove \Cref{cor:expected_grid} for sparsified grid graphs:
\begin{corollary}\label{cor:expected_grid_sparse}
 Let $d = n^c$ for any choice of $0 < c < 1$. Let $G_{n,d}$ be a random $d$-regular graph on $n$ vertices and let $H$ be a sparsified grid graph having $v$ vertices for any $v$ satisfying $v^2 \cdot d = \mathcal{O}(n)$ and $v = o(d^{1/3})$. Let $Y_H$ denote the number of induced copies of $H$ in $G_{n,d}$. Then, 
\begin{equation}
\mathbb{E}(Y_H) = \Theta\left(\left(\frac{d^{1+1/n}}{n}\right)^v \right).
\end{equation}
\end{corollary}
Hence for any $c>0$ such that $d = \Theta(n^c)$ the expectation value of $\mathbb{E}(Y_H)$ is asymptotically growing. 
\subsection{Upper bound on the variance for grid graphs}
\label{section: variance}

We currently have a good estimate of the expected number of induced subgraphs isomorphic to a square grid. However we would like to show that a large fraction of graphs includes at least one such induced graph. We can do this with the second moment method (which is really just Chebychev's inequality). \\

For each copy $H'$ of $H$ in $K_n$, we define the indicator random variable $J_{H'} = 1$ iff $H'$ is an induced subgraph of $G_{n,d}$. Then, 
\begin{equation}
Y_H = \sum_{H'} J_{H'},
\end{equation}
and
\begin{equation}
\operatorname{Var}(Y_H) = \sum_{H', H''} \operatorname{Cov}(J_{H'}, J_{H''}) = \sum_{H', H''} \left( \mathbb{E}(J_{H'} J_{H''}) - \mathbb{E}(J_{H'}) \mathbb{E}(J_{H''}) \right).
\end{equation}
We can now calculate the variance: 
\begin{lemma}\label{lemma:variance}
Let $d = n^c$ for any choice of $0 < c < 1$. Let $G_{n,d}$ be a random $d$-regular graph on $n$ vertices and let $H$ be a grid graph having $v$ vertices for any $v$ satisfying $v^2 \cdot d = \mathcal{O}(n)$, $v = o(d^{1/3})$, and $d = \omega(n^{0.5})$. Let $Y_H$ denote the number of induced copies of $H$ in $G_{n,d}$. Then, 
\begin{equation}
\operatorname{Var}(Y_H) = o\left( \mathbb{E}(Y_H)^2 \right).
\end{equation}
\end{lemma}
\noindent After proving \Cref{lemma:variance}, we extend it to the case of the sparsified grid, as defined in \Cref{subsection: sparsified grid}. 
\noindent This immediately leads to the main theorem: 
\begin{theorem}[name=Restatement of \Cref{thm:grid_subgraph}]
Let $G$ be a random $d$--regular graph on $n$ vertices, with $d= n^c$ where $0.5 < c < 1$. Then, with probability $1 - o(1)$, it contains a square grid graph on $v$ vertices, for any $v= o(n^k$), with $k = \mathsf{min}\left\{\frac{1-c}{2}, \frac{c}{3} \right\}$, as an induced subgraph.
\end{theorem}
\begin{proof}
Let $Y_{H}$ be the number of copies of $H$ in $G_{n,d}$. Observe that $v= o(n^k$), with $k = \mathsf{min}\left\{\frac{1-c}{2}, \frac{c}{3} \right\}$ satisfies $v^2 \cdot d = \mathcal{O}(n)$ and $v = o(d^{1/3})$. 

From the calculations in \Cref{cor:expected_grid} and \Cref{lemma:variance}, by applying Chebyshev's inequality,
\begin{equation}
\Pr(Y_H = 0) \leq \frac{\operatorname{Var}(Y_H)}{\mathbb{E}(Y_H)^2} = o(1).
\end{equation}
\noindent Hence, with probability $1 - o(1)$, $Y_{H}$ is non-zero, where the probability is taken over the randomness in the choice of $G_{n, d}$. This completes the proof.
\end{proof}

\noindent Now we prove \Cref{lemma:variance}. 

\begin{proof}[Proof of \Cref{lemma:variance}]

We estimate the variance of $Y_H$ by dividing it into three cases, depending on what the overlap of $H'$ and $H''$ looks like. 
\paragraph{First case (one common vertex):}
Let $H'$ and $H''$ have at most one vertex in common. 
Then, by \Cref{lemma: probability_induced_subgraph},
\begin{equation}
\label{eq: first line}
\mathbb{E}(J_{H'} J_{H''}) = \left( 1 - o(1) \right) \left( \frac{d}{n} \right)^{2e} \left(1 - \frac{d}{n} \right)^{2\binom{v}{2} - 2e} = \left(1 - o(1)\right) \mathbb{E}(J_{H'}) \mathbb{E}(J_{H''}).
\end{equation}
Now we calculate
\begin{equation}
\begin{aligned}
\sum_{\lvert V(H') \cap V(H'') \rvert \leq 1}& \left( \mathbb{E}(J_{H'} J_{H''}) - \mathbb{E}(J_{H'}) \mathbb{E}(J_{H''}) \right) \\
&\leq \sum_{\lvert V(H') \cap V(H'') \rvert \leq 1} o(\mathbb{E}(J_{H'}) \mathbb{E}(J_{H''})) \\
&\leq \sum_{\lvert V(H') \cap V(H'') \rvert \leq 1} o \left( \left( \frac{d}{n} \right)^{2e} \left(1 - \frac{d}{n} \right)^{2\binom{v}{2} - 2e} \right)
\\
&= o \left( n^{2v} \left( \frac{d}{n} \right)^{2e} \left(1 - \frac{d}{n} \right)^{2\binom{v}{2} - 2e} \right)
\\
&= o \left( \mathbb{E}(Y_H)^2 \right).
\end{aligned}
\end{equation}
The second line follows from \Cref{eq: first line}. The third line follows from \Cref{lemma: probability_induced_subgraph}. The fourth line follows from the fact that there are 
\begin{equation}
\begin{aligned}
&\left(\binom{v}{1}\binom{n-1}{v-1} \cdot \binom{n-v}{v-1} + \binom{n}{v} \cdot \binom{n-v}{v}\right)\frac{(v!)^2}{(\mathsf{aut}(H))^2}= o\big(n^{2v}\big)
\end{aligned}
\end{equation}
\noindent choices of $H'$ and $H''$ such that $\lvert V(H') \cap V(H'') \rvert \leq 1$.
\paragraph{Second case (shared induced subgraph):}
For the second case, let the intersection of $H'$ and $H''$ be a non-empty graph $F$ with $v_{F}$ vertices and $e_{F}$ edges.
 For a fixed $F$, there are at most 
\begin{equation}
\binom{v}{v_F} \cdot \binom{n-v_F}{v-v_F} \cdot \binom{n- v}{v-v_F} \cdot (v!)^2 = o\left(n^{2v - v_F}\right)
\end{equation}
\noindent choices of $H'$ and $H''$. Hence,
\begin{equation}
\begin{aligned}
\sum_{E(H') \cap E(H'') \neq \emptyset}& \left( \mathbb{E}(J_{H'} J_{H''}) - \mathbb{E}(J_{H'}) \mathbb{E}(J_{H''}) \right) \\
&\leq \sum_{E(H') \cap E(H'') \neq \emptyset} \mathbb{E}(J_{H'} J_{H''}) \\
&\leq \sum_{F \subseteq H, e_F > 0}~o\big(n^{2v - v_{F}}\big)~ \mathbb{E}(J_{H'} J_{H''}) \\
&= o\left( \sum_{F \subseteq H, e_F > 0} {n^{2v - v_{F}}} \left( \frac{d}{n} \right)^{2e - e_F} \left( 1 - \frac{d}{n} \right)^{2\binom{v}{2} - 2e + e_{F} - \binom{v_F}{2}} \right) \\
&= o\left( n^{2v} \left( \frac{d}{n} \right)^{2e} \left( 1 - \frac{d}{n} \right)^{\binom{v}{2} - 2e} \sum_{F \subseteq H, \, e_F > 0} \frac{n^{e_F - v_F}}{d^{e_F}} \left( 1 - \frac{d}{n} \right)^{e_F - \binom{v_F}{2}} \right)\\
&= o\left( \mathbb{E}(Y_H)^2 \right),
\end{aligned}
\end{equation}
when
\begin{equation}
\sum_{F \subseteq H, \, e_F > 0} \frac{n^{e_F - v_F}}{d^{e_F}} \left( 1 - \frac{d}{n} \right)^{e_F - \binom{v_F}{2}} = o(1).
\end{equation}
The fourth line follows from \Cref{expected_value}. Now, note that the number of subgraphs of $F$ is 
\begin{equation}
2^{e_F} = n^{{e_{F}}/{\log n}}.
\end{equation}
\noindent Now,
\begin{equation}
\sum_{F \subseteq H, \, e_F > 0} \frac{n^{e_F - v_F}}{d^{e_F}} \left( 1 - \frac{d}{n} \right)^{e_F - \binom{v_F}{2}} = n^{{e_{F}}/{\log n}} \cdot \frac{n^{e_F - v_F}}{d^{e_F}} \left( 1 - \frac{d}{n} \right)^{e_F - \binom{v_F}{2}} = o(1),
\end{equation}
\noindent whenever
\begin{equation}
\frac{n^{e_F - v_F}}{d^{e_F}} = o\left( \frac{1}{n^{{e_{F}}/{\log n}}}\right).
\end{equation}
This estimation holds whenever
\begin{equation}
\label{equation: degree requirement}
d > n^{1 + 1/\log n -v_{F}/e_{F}}.
\end{equation}
\noindent By the definition of the density $m(H)$, one sufficient condition for \Cref{equation: degree requirement} to hold is
\begin{equation}
d > n^{1 + 1/\log n - 1/m(H)}.
\end{equation}
\noindent Note that the other term
\begin{equation}
\left( 1 - \frac{d}{n} \right)^{e_F - \binom{v_F}{2}} = o(1).
\end{equation}
\noindent Choosing $H$ to be a grid graph, it is easy to see that $m(H)$ is at most $2$. Therefore
\begin{equation}
d = \omega(n^{0.5}).
\end{equation}
makes the overall term negligible.
\paragraph{Third case (empty shared induced subgraph):}
Now, we consider the case when $H'$ and $H''$ have $t$ vertices in common but no edges in common. There are 
\begin{equation}
\binom{v}{t} \cdot \binom{n-t}{v-t} \cdot  \binom{n-v }{v-t} \cdot \left(\frac{(v-t)!}{\mathsf{aut}(H)}\right)^2  = o(n^{2v - t}).
\end{equation}
\noindent such cases. The second line follows from the following estimations:
\begin{equation}
\binom{n}{t} = o(n^t), ~~ \binom{n-t}{v-t} = o(n^{v-t}), ~~ \binom{n-v}{v-t} = o(n^{v-t}).
\end{equation}
Again by \Cref{lemma: probability_induced_subgraph}, we have that
\begin{equation}
\label{eq: subtract t/2}
\mathbb{E}(J_{H'} J_{H''}) = \left(1 + o(1)\right) \left( \frac{d}{n} \right)^{2e} \left( 1 - \frac{d}{n} \right)^{2\binom{v}{2} - 2e - \binom{t}{2}}.
\end{equation}
\noindent In \Cref{eq: subtract t/2}, we subtract $\binom{t}{2}$ edges from the exponent of the term in the extreme right because we are counting them twice --- once in the complement set of $H'$ and again in the complement set of $H''$.
Hence,
\begin{equation}
\begin{aligned}
\sum_{\substack{{E}(H') \cap {E}(H'') = \emptyset \\ \lvert V(H') \cap V(H'') \rvert \leq t}}& \left( \mathbb{E}(J_{H'} J_{H''}) - \mathbb{E}(J_{H'}) \mathbb{E}(J_{H''}) \right) \\
&\leq \sum_{\substack{{E}(H') \cap {E}(H'') = \emptyset \\ \lvert V(H') \cap V(H'') \rvert \leq t}} \mathbb{E}(J_{H'} J_{H''}) \\
&= o\left( \sum_{t \geq 2} n^{2v - t} \left( \frac{d}{n} \right)^{2e} \left(1 - \frac{d}{n}\right)^{2\binom{v}{2} - 2e - \binom{t}{2}} \right) \\
&= o\left( n^{2v} \left( \frac{d}{n} \right)^{2e} \left(1 - \frac{d}{n}\right)^{2\binom{v}{2} - 2e} \sum_{t \geq 2} n^{-t} \left(1 - \frac{d}{n}\right)^{-\binom{t}{2}} \right) \\
&= o\left( \mathbb{E}(Y_H)^2 \right).
\end{aligned}
\end{equation}
\noindent The last line follows from the fact that
\begin{equation}
\sum_{t \geq 2} n^{-t} \left(1 - \frac{d}{n}\right)^{-\binom{t}{2}} = o(1).
\end{equation}
\noindent Therefore, putting everything together we have
\begin{equation}
\operatorname{Var}(Y_H) = o\left( \mathbb{E}(Y_H)^2 \right),
\end{equation}
which is the lemma statement
\end{proof}

\begin{corollary}\label{lemma:variance2}
Let $G_{n,d}$ be a random $d$-regular graph on $n$ vertices and let $H$ be a sparsified square grid graph having $v$ vertices for any $v$ satisfying $v^2 \cdot d = \mathcal{O}(n)$, $v = o(d^{1/3})$, and $d = n^c$ for any constant $0 < c < 1$. Let $Y_H$ denote the number of induced copies of $H$ in $G_{n,d}$. Then, 
\begin{equation}
\operatorname{Var}(Y_H) = o\left( \mathbb{E}(Y_H)^2 \right).
\end{equation}
\end{corollary}
\begin{proof}
The proof is the same as that of \Cref{lemma:variance} with the observation that in the second case, if $H$ is the sparsified square grid graph, the local density is bounded as
\begin{equation}
m(H) \leq \frac{\sqrt{v}}{\sqrt{v}-1}.
\end{equation}
\noindent Hence
\begin{equation}
n^{1 + 1/\log n - 1/m(H)} \leq n^{1/m + 1/\sqrt{v}} < n^c = d,
\end{equation}
for any constant $0 < c < 1$.
\end{proof}

\section{Geometric entanglement of high degree graph states}
\label{sec: high degree}

In this section we investigate the computational complexity of random regular graph states of high degree, i.e. $d  = cn $ with $c\in (0,1)$. We will use \emph{uniformly} random graphs as a proxy for graphs of high degree. We suspect that these graphs are universal with high probability. However, the proofs given in the previous section explicitly break down in the regime where $d = cn$. More strongly, it is known that in this regime one can not find large (much larger than $\log(n)$ sized) non-trivial \emph{induced subgraphs} with more than negligible probability~\cite{rucinski1987induced}. This leads one to suspect the contrary of our earlier assertion, namely that graph states of high degree are almost never universal resources. Indeed this is the case for \emph{Haar random states} \cite{gross_most_2009}, which with high probability have geometric entanglement so high that MBQC measurements can be effectively simulated by coin flips. In this section we show that this is not the case for random graph states, by providing (almost matching) upper and lower bounds on the (expected) geometric entanglement. This is not proof positive of universality, but at least we avoid one known barrier. At the end of the section we discuss extensions to this simulation barrier specific to stabilizer states, conjecturing that this too can be avoided.

\subsection{Lower bound on geometric entanglement}
In this section we will prove \Cref{thm:geom_ent_lower_stab} and \Cref{cor:geom_ent_lower_graph}. We begin by extending a nice trick from compressed sensing (see e.g. \cite[Lemma 4.4.1]{vershynin2018high}) on approximating extremal singular values of matrices through epsilon nets (we found essentially this argument in \cite{tomioka2014spectral} but it is probably folklore in the tensor community). 

\begin{lemma}\label{lem:eps_net_bound}
Consider an $n$ qubit state $\ket{\psi}$ and let $A_{\epsilon/n}$ be an $(\ln(3/2)/n)$-net of the set of single qubit states. We then have that
\begin{equation}
\label{eq:product maximization alpha*}
\max_{\ket{\alpha}\in \mathrm{PROD}} |\langle \alpha|\psi\rangle|^2 \leq  2 \max_{\ket{\beta}\in A_{\ln(3/2)/n}\tn{n}} |\langle \beta|\psi\rangle|^2.
\end{equation}
\end{lemma}
\begin{proof}
Consider a state $\ket{\alpha^*}$ such that $ \max_{\ket{\alpha}\in \mathrm{PROD}}|\langle \alpha|\psi\rangle|^2 = |\langle \alpha^*|\psi\rangle|^2$. Writing $\ket{\alpha^*} = \bigotimes_{i=1}^n \ket{\alpha^*_i}$ we choose for each $\ket{\alpha^*_i}$ a state $\ket{\beta_i}\in A_{\ln(3/2)/n}$ such that $\norm{\ket{\alpha^*_i} -\ket{\beta_i} }\leq \ln(3/2)/n$. We can write
\begin{align}
|\langle \alpha^*|\psi\rangle| = |\langle \bigotimes_{i=1}^n (\bra{\beta_i} + \bra{\alpha^*_i} -  \bra{\beta_i})|\psi\rangle|\leq |\langle \beta|\psi\rangle| + \sum_{i=1}^n \binom{n}{i}(\ln(3/2)/n)^i |\langle \alpha^*|\psi\rangle|,
\end{align}
using the fact that $\ket{\alpha^*}$ maximizes the overlap with $\ket \psi$ over normalized product states so that $|\braket{\alpha^*}{\psi} - \braket{\beta}{\psi} | \le \norm{ \ket \alpha - \ket \beta} |\braket{\alpha^*}{\psi}|$. 
The bound then follows from applying this fact for every tensor factor and $\norm{\ket{\alpha_i^*}-\ket{\beta_i}} \leq \ln(3/2)/n$. Now note that
\begin{equation}
 \sum_{i=1}^n \binom{n}{i}(\ln(3/2)/n)^i = (1+ \ln(3/2)/n)^n - 1 \leq e^{\ln(3/2)} -1 =\frac{1}{2}.
\end{equation}
We can thus invert the relation above to get the lemma statement.
\end{proof}
This lemma allows us to get a multiplicative approximation to the maximum overlap using a relatively weak (and thus small) epsilon net for the product states.

Recall that we are trying to lower-bound the geometric entanglement. 
This is equivalent to upper-bounding the maximization over product states. 
The probability that the geometric entanglement is small is equal to the probability that the maximization of the overlap is large. 
Now, we upper bound this probability.
Combining the above argument with the union bound and the formula for average stabilizer states \Cref{eq:average_stab},   we obtain:
\begin{theorem}[restatement of \Cref{thm:geom_ent_lower_stab}]
Choose a stabilizer state $\ket S$ uniformly at random. There exists a constant $c$ such that
\begin{equation}
\mathbb{P}\bigg[E_g(\ket{S}) \leq n - c\sqrt{n}\log(n) \bigg]\leq O(2^{-\sqrt{n}}).
\end{equation}
\end{theorem}
\begin{proof}
We begin by noting that 
\begin{multline}
\mathbb{P}\Big[E_g(\ket{S}) \leq n-\delta\Big] = 
\mb P\left[  -\log\big(\max_{\alpha \in \mathrm{PROD}_n}|\braket{\alpha}{\psi}|^2\big) \le n - \delta\right]\\
=
\mathbb{P}\Big[\max_{\ket{\alpha}\in \mathrm{PROD}} |\langle \alpha|S\rangle|^2 \ge 2^{-n+\delta}\Big] \le  \mathbb{P}\Big[\max_{\ket{\beta}\in A_{1/n}\tn{n}} |\langle \beta|S\rangle|^2\geq 2^{-n+\delta-1} \Big],
\end{multline}
where $A_{\ln(3/2)/n}$ is a $\ln(3/2)/n$-net for the set of single qubit states. We know that there exists such a net of size $|A_{\ln(3/2)/n}|\leq (5n/\ln(3/2))$~\cite{hayden_randomizing_2004}. Using the union bound and Markov's inequality we can upper bound this latter quantity as
\begin{equation}
\mathbb{P}\Big[\max_{\ket{\beta}\in A_{\ln(3/2)/n}\tn{n}} |\langle \beta|\psi\rangle|^2\leq 2^{-n+\delta-1} \Big]\leq 2^{n-\delta+1} \mathbb{E}_S\bigg[\sum_{\ket{\beta}\in A_{\ln(3/2)/n}\tn{n}} |\langle \beta|S\rangle|^{2t}\bigg ]^{1/t} 
\end{equation}
for some integer $t>1$ (we will specify this later)
Next we use the concavity of $x^{1/t}$ and  the duality formula described in \Cref{eq:average_stab} to obtain
\begin{equation}
2^{n-\delta-1} \mathbb{E}_S\bigg[\sum_{\ket{\beta}\in A_{\ln(3/2)/n}\tn{n}}\hspace{-1.5em} |\langle \beta|S\rangle|^{2t}\bigg ]^{1/t} \!\!\!\!\leq 2^{n-\delta-1} |A_{\ln(3/2)/n}|^{n/t} 2^{-n/t} (2^{t-1}+1)^{1/t} \bigg[\prod_{i=0}^{t-2} \frac{(2^i+1)}{(2^n\!+\!2^i)}\bigg]^{1/t}\hspace{-1em}.
\end{equation}
Choosing $t =\sqrt{n}$ and working out we obtain
\begin{equation}
2^{n-\delta+1} \mathbb{E}_S\bigg[\sum_{\ket{\beta}\in A_{\ln(3/2)/n}\tn{n}}|\langle \beta|S\rangle|^{2t}\bigg ]^{1/t} \leq 5^{\sqrt{n}} 2^{-\delta} 2^{\log(n) \sqrt{n}} 2^{\sqrt{n}} .
\end{equation}
Thus setting $\delta  =  2 \log(n) \sqrt{n}$ we obtain what we set out to prove.
\end{proof}
The above statement holds for uniformly random stabilizer states. We can lift it to a statement about graph states by noting that the probability that a uniformly random stabilizer state has full support on the computational basis states is bounded from below by a constant $C$. This is a standard fact, but we prove it here for completeness:
\begin{corollary}[restatement of \Cref{cor:geom_ent_lower_graph}]
Choose a graph state $\ket G$ uniformly at random. There exists a constant $c$ such that
\begin{equation}
\mathbb{P}\bigg[E_g(\ket{G}) \leq n - c\sqrt{n}\log(n) \bigg]\leq O(2^{-\sqrt{n}}).
\end{equation}
\end{corollary}
\begin{proof}
Consider the stabilizer states of full support:
\begin{equation}
\ket{S_{U,b}} := 2^{-n/2}\sum_{x\in \{0,1\}^n} (-1)^{x^TUx}\, i^{x^Tb} \ket{x}
\end{equation}
with $U$ a binary upper triangular matrix and $b\in \mathbb{Z}_4^n$. We will argue that all pairs $U,b$ correspond to different quantum states. For two such states we can compute the inner product:
\begin{align}
\braket{S_{V,a}}{S_{U,b}} = 2^{-n}\sum_{x\in \{0,1\}^n} (-1)^{x^T(U+V)x} i^{x^T(a+b)}.
\end{align}
The matrix $U+V \mod 2$ is again upper triangular, and hence a theorem due to Dickson \cite[Ch. 15 Thm. 2]{macwilliams1977theory} tells us that there exists an invertible binary matrix $P$ such that $P(U+V)P^{-1} =  D$ with
\begin{equation}
x^T Dx = \sum_{i=1}^{\mathrm{rank}(U+V)/2}{x}_{2i-1} {x}_{2i}.
\end{equation}
Note that the rank of $U+V$ is always even, so this is sensible. Absorbing the map $P$ into the summation, we can write:
\begin{equation}
\braket{S_{V,a}}{S_{U,b}} = 2^{-n}\sum_{x\in \{0,1\}^n} (-1)^{x^TDx} i^{(Px)^T (a+b)}.
\end{equation}
In order for this overlap to be one we must have 
\begin{equation}
(-1)^{x^TDx} i^{(Px)^T (a+b)} = 1 ,
\end{equation}
for all $x\in \{0,1\}^n$. This immediately implies that $i^{(Px)^T (a+b)} =\pm 1$ and hence that $a+b = 2z$ for some $z\in\{0,1\}^n $. Furthermore we must have 
$x^TDx + x^T  (P^T z)= 0$ for all $x\in \{0,1\}^n$. Setting $P^T z = z'$ we can write this as
\begin{equation}
0= \sum_{i=0}^{r/2} x_{2i}x_{2i+1} + x_{2i}z'_{2i} +x_{2i+1} z'_{2i+1}+ \sum_{i=r/2+1}^{n /2} x_{2i}z'_{2i} +x_{2i+1} z'_{2i+1},
\end{equation}
where $r$ is the rank of $D = U+V$. Clearly $z'_{2i} = z'_{2i-1}  =0$ for $i> r/2$. Moreover if $r>1$ it is easy to see (by explicit enumeration) that for any choice of $z'_0,z'_1$ one of the evaluations of $x_0x_1 +x_0z'_0 +x_1z'_1$ is always nonzero. Hence we must have $r=0$, which implies $D=0$. This also implies at $a+b=0$, since $P$ is invertible. Hence we must have that $U=V$ and $a=b$, which is what we set out to argue. With this it is clear that the set $\{\ket{S_{U,b}}\}_{A,b}$ is of size $2^{n(n-1)/2} \times 2^{2n}$. Comparing this to the total number of stabilizer states $2^n \prod_{i=1}^{n} (2^{i} +1)$ we see that 
\begin{equation}
2^{\frac{n(n-1)}{2} + 2n} \bigg[ 2^n \prod_{i=1}^{n} (2^{i} +1)\bigg]^{-1} =2^{\frac{n(n-1)}{2} + 2n -n - \frac{n(n+1)}{2}} \prod_{i=1}^n (1 +2^{-i})^{-1} \geq \prod_{i=1}^\infty (1 +2^{-i})^{-1} = C
\end{equation}
where $C^{-1}\approx 2.3842$ (by numerical evaluation).
Now we can prove the actual corollary statement by noting that (1) the geometric entanglement of $\ket{S}_{U,b}$ is independent of $b$, and (2) that the the set $\{\ket{S}_{U,b}\}_{b}$ contains exactly one graph state (namely $b=0$). From this we can see that
\begin{equation}
\mathbb{P}_{\ket{G}}\big[ E_g(\ket{G})\leq n-\delta \big] = \mathbb{P}_{\ket{S_{U,b}}}\big[ E_g(\ket{S_{U,b}})\leq n-\delta \big],
\end{equation}
where in the RHS we choose $U,b$ uniformly at random.
Since conditioning on inclusion in a subset under a uniform distribution yields a uniform distribution on that subset we can say that 
\begin{equation}
\mathbb{P}_{\ket{S_{U,b}}}\big[ E_g(\ket{S_{U,b}})\leq n-\delta \big]\leq C^{-1} \mathbb{P}_{\ket{S}}\big[ E_g(\ket{S})\leq n-\delta \big],
\end{equation}
which is what we wanted to show.
\end{proof}

\subsection{Upper bound on geometric entanglement}
The main goal of this section is to provide a proof of the upper bound in \Cref{thm:geom_ent_upper}. The proof of this statement is substantially more difficult than the associated lower bound in \Cref{cor:geom_ent_lower_graph}. We begin by proving a lemma that exactly characterizes the maximal overlap of a random graph state with the set of real stabilizer product states: $R_n = \{\ket{0},\ket{1},\ket{+},\ket{-}\}\tn{n}$. 
This is a relatively small subset of all product states, 
and therefore gives a lower bound on the maximal overlap, which translates into an upper bound for the expected geometric entanglement since the overlap enters with a minus sign. 
\begin{lemma}\label{lem:geom_ent_upper}
Consider the set of real stabilizer product states: $R_n = \{\ket{0},\ket{1},\ket{+},\ket{-}\}\tn{n}$. We then have that 
\begin{equation}
\mathbb{E}_G (E_g(\ket{G}))\leq \mathbb{E}_G \left[- \log(\max_{\ket{s}\in R_n}|\bra{G}s\rangle|^2)\right] =  n - \mathbb{E}_G\left[\max_{S\subseteq [n]}\left( |S| - \Rank{A_G[S]}\right) \right],    
\end{equation}
where $A_G[S] \equiv A_G[S,S]$ is the adjacency matrix of $G$ restricted to the index set $S$. 
\end{lemma}
\begin{proof}
The first inequality is obvious from the definition of the geometric entanglement $E_g$. Now note that we can specify any state $\ket s \in R_n$ by a set $S\subset [n]$ indicating the Hadamard basis part of the state and two bitstrings $x_S\in \{0,1\}^{|S|}$ and $y_{\bar{S}} \in \{0,1\}^{n- |S|}$ encoding the phase and computational basis states in $S,\bar{S}$ so that $\ket s = 2^{-|S|/2} \sum_{z \in \bin^{|S|}} (-1)^{x_S^T z}\ket{z} \ket{y_{\bar S}}$. 
To ease notation, let us assume wlog.\ that $S = \{1, \ldots, |S|\}$, i.e., the submatrix $A_G[S]$ is just the top-left corner of $A_G$. Then 
\begin{equation}
\hspace{-.9em}\max_{\ket{s}\in R_n}|\bra{G}s\rangle|^2 = 2^{-n - |S|}\max_{S\subset[n]}\max_{x_S\in \bin^{|S|}}\max_{y_{\bar{S}} \in \bin^{n-|S|}} \left|\sum_{z\in \bin^{|S|}} (-1)^{(z,y_{\bar{S}})^T U_G(z,y_{\bar{S}}) + z^Tx_S}\right|^2\hspace{-0.5em},
\end{equation}
where we recall that $U_G$ is the adjacency matrix of $G$ with the lower triangular part set to zero. Splitting the matrix $U_G$ into submatrices ${U}_{SS},{U}_{S\bar{S}},{U}_{\bar{S}S}, {U}_{\bar{S}\bar{S}}$ we can write the phase inside the summation as 
\begin{equation}
(z,y_{\bar{S}})^TU(z,y_{\bar{S}}) + x_S^Tz = z^T{U}_{SS} z+ \big(x_S^T +  y_{\bar{S}}^T {U}_{S\bar{S}}^T +y_{\bar{S}}^T{U}_{\bar{S}S}\big)z  + y_{\bar{S}}^T{U}_{\bar{S}\bar{S}}y_{\bar{S}}.
\end{equation}
The final term in this equation contributes only a global phase, while the middle one can be absorbed in the maximization over $x_S$. Hence we get
\begin{equation}
\max_{\ket{s}\in R_n}|\bra{G}s\rangle|^2=2^{-n - |S|} \max_{S\subset[n]}\max_{x_S\in \bin^{|S|}}\bigg|\sum_{z\in \bin^{|S|}} (-1)^{z^T{U}_{SS} z+ x_S^Tz}\bigg|^2.
\end{equation}
Next we use a theorem due to Dickson \cite[Ch. 15 Thm. 2]{macwilliams1977theory}, which tells us that there exists an invertible binary matrix $P_{SS}$ such that $P_{SS}U_{SS}P^{-1}_{SS} =  D_{SS}$ with
\begin{equation}
z^T D_{SS}z = \sum_{i=1}^{\mathrm{rank}(U_{SS})/2}{z}_{2i-1} {z}_{2i}.
\end{equation}
Note that the rank of $U_{SS}$ is always even, so this equation makes sense.
Absorbing $P_{SS}$ in the sum over $z$ and subsequently into the maximisation over $x_S$ we obtain
\begin{equation}
\max_{\ket{s}\in R_n}|\bra{G}s\rangle|^2 =\max_{S\subset[n]}\max_{x_S\in \{0,1\}^{|S|}}\bigg|\sum_{a_S\in \{0,1\}^S} (-1)^{z^T{D}_{SS} z+ x_S^Tz}\bigg|^2.
\end{equation}
Using the definition of $D_{SS}$ we see that this equation factorises, and we obtain
\begin{equation}
\max_{\ket{s}\in R_n}|\bra{G}s\rangle|^2=2^{-n - |S|}\max_{S\subset[n]}\bigg(\max_{y_1,y_2\in\{0,1\} } \bigg|\sum_{a_1,a_2\in \{0,1\}} (-1)^{a_1a_2 + y_1a_1 +y_2 a_2}\bigg|^2\bigg)^{\mathrm{rank}(U_{SS})/2}.
\end{equation}
The inner maximisation can easily be solved to obtain
\begin{equation}
\max_{\ket{s}\in R_n}|\bra{G}s\rangle|^2=2^{-n - |S|} \,2^{\mathrm{rank}(U_{SS})}.
\end{equation}
Noting that $\mathrm{rank}(U_{SS}) = \mathrm{rank}(A_G[S])$ we obtain the lemma statement. 
\end{proof}

This means that the expected geometric entanglement can be controlled by the average rank of submatrices of random adjacency matrices. To determine this, we will need two facts from probability theory and classical coding theory: 

\begin{fact}[Bonferroni inequalities]\label{fact:bonferroni}
Let $\{E_i\}_{i\in \Omega}$ be a countable set of events, then:
\begin{align}
\mathbb{P}\bigg(\bigcup_{i\in \Omega} E_i\bigg)&\leq \sum_{i\in \Omega}  \mathbb{P}\big( E_i\big),\tag{union bound}\\
\mathbb{P}\bigg(\bigcup_{i\in \Omega} E_i\bigg)&\geq \sum_{i\in \Omega}  \mathbb{P}\big( E_i\big) - \frac{1}{2}\sum_{i,j\in \Omega,\;i\neq j}\mathbb{P}\big( E_i\cup E_j\big). \tag{Bonferroni}
\end{align}
\end{fact}
Our goal will be to use the second inequality to lower bound the probability that any submatrix $A_G[S]$ of a random adjacency matrix $A_G$ has excessive rank deficit. We can characterise the distribution of this rank for a fixed $S$ exactly:
\begin{fact}[random adjacency matrices]\label{fact:rand_adj_rank}
Consider the set of $n\times n$ binary symmetric matrices with zeros on the diagonal (i.e. adjacency matrices). Choosing a matrix $A$ uniformly from this set we have $\mathbb{P}\big(\mathrm{Rk}(A_G) = 2h+1\big) = 0$) and:
\begin{equation}\label{eq:prob_rank}
\mathbb{P}\big(\mathrm{Rk}(A_G) = 2h\big) = 2^{-n^2/2 + n/2}\prod_{i=1}^{h} \frac{2^{2i-2}}{2^{2i}-1} \prod_{i=0}^{2h-1}(2^{n-i} - 1),
\end{equation}
with $h \in [0,\lfloor n/2\rfloor]$.
\end{fact}
The formula above, found in \cite[Ch.15 Thm.2]{macwilliams1977theory}\footnote{Note however that this equation in \cite{macwilliams1977theory} contains a typo, we provide the correct formula.}, can be both upper and lower bounded by a Gaussian. We have the following lemma, which follows from a straightforward calculation. 
\begin{lemma}\label{lem:gaussian_bound}
Consider a uniformly random adjacency matrix $A$ as in \Cref{fact:rand_adj_rank}. We have the following approximation:
\begin{equation}
\frac{e^{-2}}{4}2^{-\frac{(n-2h)^2}{2}} 2^{\frac{(n-2h)}{2}} \leq \mathbb{P}\big(\mathrm{Rk}(A) = 2h\big)\leq e^{2/3} 2^{-\frac{(n-2h)^2}{2}} 2^{\frac{(n-2h)}{2}},
\end{equation}
with $h \in [0,\lfloor n/2\rfloor]$.
\end{lemma}
This immediately translates into a tail bound on the \emph{rank deficiency} of a submatrix $A_G[S]$ for a fixed $S$.
\begin{lemma}\label{lem:event_prob}
Consider a uniformly random $n\times n$ adjacency matrix $A$ and a set of indices $S\subseteq [n]$ and define the event $E_S(t) = \{ |S| - \mathrm{Rk}(A)\geq t\}$ with $ t\in [0, |S|]$. We have that
\begin{equation}
\mathbb{P}\big(E_S(t)\big)\geq \frac{e^{-2}}{4} 2^{-\frac{t^2}{2} - \frac{t}{2}}.
\end{equation}
\end{lemma}
This allows us to control the first term in the Bonferroni inequality. To characterize the second term will we need to do substantially more work. First we note that the rank of two submatrices $A[S],A[S']$ is independent conditioned on the rank of the intersection matrix $A[S\cap S']$.

\begin{lemma}\label{lem:independence intersection}
Consider a uniformly random $n\times n$ adjacency matrix $A$ and two sets of indices $S,S'\subseteq [n]$, with $S\cap S' \eqqcolon I$. Conditioned on the rank of the intersection $(\,\Rank{A[I]}\,)$ the ranks of $A[S]$ and $A[S']$ are independent, i.e.
\begin{align}
&\mathbb{P}\big(\Rank{A[S]}= i, \Rank{A[S']} = i'\big)\\
 &\hspace{2em}=\sum_{j=0}^{\lfloor|I|/2\rfloor}\mathbb{P}\big(\Rank{A[S]} = i\; \big|\; \Rank{A[I]} = j \big)\mathbb{P}\big(\Rank{A[S']} = i' \;\big|\; \Rank{A[I]} = j \big)\mathbb{P}\big(\Rank{A[I]} = j \big).\notag
\end{align}
\end{lemma}
\begin{proof}
Certainly $A[S]$ and $A[S']$ are independent conditioned on the intersection $A[I]$, since the matrix elements $A[S]$ and $A[S']$ outside of the intersection are independent. 
It is thus sufficient to establish that the rank of $A[S]$ only depends on the rank of $A[I]$ (and similarly for $A[S']$). Consider two matrices $B_I,B'_I$ with $\Rank{B_I} = \Rank{B_I'}$. 
From Dickson's theorem we know there exists an invertible matrix $R_I$ such that $B_I' = R_I B_I R_I^T$. 
Now consider the induced distributions $\mathbb{P}\big(A[S]\,|\, A[I] = B_I \big)$ and $\mathbb{P}\big(A[S'] \,|\, A[I] = B'_I\big)$. We have $ \mathbb{P}\big(A[S] \,|\,  A[I] = B_I\big) = \mathbb{P}\big((\mathds{1}\oplus R_I^{-1}) A[S'](\mathds{1}\oplus {R_I^{T}}^{-1}) \,| \,A[I] = B_I\big)$. Clearly $\Rank{(\mathds{1}\oplus R_I^{-1}) A[S'](\mathds{1}\oplus {R_I^{T}}^{-1})} = \Rank{A[S']}$, which gives us the desired result. 
\end{proof}
With the above we can control the joint probability of the rank of two submatrices $A[S],A[S']$ in terms of only their intersection. Next we will show that if $|S|$ is much larger than $|I|$ then the rank of $A[S]$, conditioned on the rank of $A[I]$ is close to maximal with high probability (provided that the rank of $A[I]$ is not too small). We do this by a reduction of the problem to an infinite dimensional Markov chain together with precise bounds on its convergence rate.

\begin{lemma}\label{lem:cond_upper_bound}
Consider a uniformly random $n\times n$ adjacency matrix $A$ and two sets of indices $S,I\subseteq [n]$, with $I\subseteq S$. Also consider integers $i,j$ such that $|S|-i, |I|-j$ are even. 
There exists a constant $\rho<1-10^{-5}$ such that 
\begin{equation}\label{eq:lem_cond_upper}
\mathbb{P}\big( \Rank{A[S]} \leq |S|-i \;\big|\;  \Rank{A[I]} = |I| -j\big) \leq e^{2/3} 2^{-\frac{(i)^2}{2}- \frac{(i)}{2}} +7\cdot 2^{j} \rho^{|S|-|I|},
\end{equation} 
for $i \geq 0$ and $\alpha >1$.
\end{lemma}
\begin{proof}
The main strategy of this proof will be to rewrite the LHS of Eq. \ref{eq:lem_cond_upper} in terms of the convergence properties of a (formally infinite) Markov chain $P$ and then use classical Markov chain bounding techniques (in particular the drift and minorization method \cite{meyn1994computable,rosenthal1995minorization}) to provide bounds on these convergence properties.

To construct the Markov chain, consider an $m\times m$ symmetric boolean matrix $A$ of rank $r$ (with zeros on the diagonal).
We will now symmetrically add a vector $v \in \bin^m$ to the rows and columns. Adding the column first, the probability that $\Rank{\begin{pmatrix} v & A \end{pmatrix}} = r$ is $2^{r-m}$ and the probability that $\Rank{\begin{pmatrix} v &A \end{pmatrix}} = r+1$ is $1-2^{r-m}$. In the second case it is clear that $\Rank{\begin{pmatrix}0 &  v^T \\v & A \end{pmatrix}} = r+2$ since column and row rank are always the same. 
In the first case we note that there exists an $x$ s.t. $Ax = v$. This immediately implies that
\begin{equation}
\begin{pmatrix} v^T \\ A \end{pmatrix}x = \begin{pmatrix} v^Tx \\ v \end{pmatrix} = \begin{pmatrix}0 \\ v \end{pmatrix},
\end{equation}
because $v^T x = x^T A x = 0$ by the symmetry of $A$ and the fact that we are working over the field $\mb F_2$. This implies that $\Rank{\begin{pmatrix}0 &  v^T \\v & A \end{pmatrix}} = r$ (this elegant argument is due to Sloane and MacWilliams~\cite[Ch. 15 Lem. 3]{macwilliams1977theory}). 
The sequential adding of random $k$ columns (and rows) to a matrix $A$ of rank $r$, and considering their rank, thus induces a sequence of random variables $R_0 = r, R_1,\ldots , R_k$. From the above discussion, this sequence is a (time dependent) Markov chain with transitions $r \xrightarrow{2^{r-m}} r$, $r \xrightarrow{1-2^{r-m}} r+2$ (where $m$ takes the role of time). 

 Changing variables from the rank $R_k$ to the rank deficiency $D_k = m+k-R_k $ we obtain another Markov chain who's transition probabilities no longer depend on the ambient matrix dimension $m$ (it is now homogeneous and formally infinite dimensional). The associated Markov generator is given explicitly as
\begin{equation}
M(j,i)\coloneqq \mathbb{P}\big(D_{k+1} =j \big| D_k = i\big)  = 
\begin{cases} 
&2^{-i} \hspace{1em}\text{if} \hspace{1em} j=i+1\\
&1-2^{-i}\hspace{1em}\text{if} \hspace{1em} j=i-1\\
&0\hspace{1em}\text{if} \hspace{1em} |j-i|>1,
\end{cases}
\end{equation}
for $i,j \in \mathbb{N}$. This Markov chain is irreducible, but is periodic with period $2$. Hence it is natural to consider $Q = M^2$, which will be aperiodic, but decomposes into even and odd irreducible aperiodic subchains. We will now bound the convergence of the Markov chain $Q$ with initial state $e_r$ (the unit vector with $1$ on the $r$'th position). We will treat the even subchain in detail (with the odd subchain being analogous), so assume that $r$ is even. The stationary state $\pi$ of $M^2$ on the even subspace can be found by appropriately taking the limit of \Cref{eq:prob_rank} to $n\to \infty$:
\begin{equation}
\pi: \mathbb{N}/2 \to \mathbb{R}: i \to C^{-1} 2^{-\frac{i^2}{2} +\frac{i}{2}} \prod_{t = \frac{i}{2}+1}^{\infty} \frac{2^{2t -i}}{2^{2i-t} - 1}\prod_{t = i+1}^{\infty} (1-2^{-t}),
\end{equation}
where $C\geq 1$ is some appropriate normalization.
The above makes intuitive sense, because it corresponds to the situation where $|S|$ is much larger than $|I|$, and thus the rank distribution of $A[|S|]$ stops depending on $I$. One can also explicitly verify that $\pi$ is an eigenvector of $M^2$ with eigenvalue $1$. We will use the drift and minorization method (see \Cref{thm:markov_conv}) to bound convergence to this distribution. For this bounding method we need to provide a drift function $V:\mathbb{N}/2\to \mathbb{R}$ and a small set $C\subset \mathbb{N}/2$. We will choose $V(i) =2^i $ and $C = \{0,2\}$. It is tedious but straightforward to check that $P^2$ satisfies the conditions of \Cref{thm:markov_conv} with parameters $\lambda = 0.55, b = 2, \delta = 0.2, d = 9, r= 0.001$ (all the difficulty lies in choosing the parameters). This bounds the convergence of the Markov chain in total variation distance as 
\begin{equation}
\norm{\pi - M^{2k}e_r}_{TV} \leq \rho_1^{2k} + \rho_2^{2k} \big(6 + 2^{r}\big)\leq 7 \rho^{k} 2^{r},
\end{equation}
with $\rho = \max\{\rho_1,\rho_2\}$ and $\rho_1 < 1-10^{-5}, \rho_2< 1- 10^{-3}$. We obtain the same expression for the odd subchain. \\

Mapping back to our original question, we see that
\begin{multline}
\mathbb{P}\left( \Rank{A[S]} \leq |S|-i \;\big|\;  \Rank{A[I]} = |I| -j\right) \\=  \mathbb{P}\left(D_{|S|-|I|} \geq i\;\big|\;\Rank{A[I]} = |I| -j\right)
\leq \sum_{t= i}^{\infty} \left(M^{|S|-|I|} e_{|I| -j }\right)_{t}, 
\end{multline}
recall that $e_i$ is $i$-th unit vector on $\mb R^{\mb N}$ and $(\cdot)_t$ denotes the $t$-the element. Using the triangle inequality on $(M^{|S|-|I|} e_{|I| -j} - \pi) +\pi$ we obtain
\begin{multline}
\mathbb{P}\big( \Rank{A[S]} \leq |S|-i \;\big|\;  \Rank{A[I]}= |I| -j\big) ]\leq 7 \rho^{|S|-|I|} 2^{j} + \sum_{t = i}^{\infty} \pi_t = 7 \rho^{|S|-|I|} 2^{j} + 2\, 2^{-\frac{i^2}{2}+ \frac{i}{2}},\notag
\end{multline}
using the definition of the distribution $\pi$ and the basic upper bound $\prod_{t = \frac{i}{2}+1}^{\infty} \frac{2^{2t -i}}{2^{2i-t}}\leq 2 $.

\end{proof}

With this lemma under our belt it is finally time to prove the main theorem of this section. 
\begin{theorem}[Restatement of \Cref{thm:geom_ent_upper}]
Choose a graph state $\ket{G}$ on $n$ qubits uniformly at random, where we assume $n = k^2$ for some integer $k$.  We have
\begin{equation}
\mathbb{E}_G \big(E_g(\ket{G}\big) \leq n - \Omega(n^{1/4}/\log(n)).
\end{equation}
\end{theorem}
\begin{proof}
We begin by upper-bounding the geometric entanglement of a graph state $\ket{G}$ in terms of the maximal rank deficiency of the principal submatrices of the adjacency matrix of $G$. \Cref{lem:geom_ent_upper} tells us that
\begin{equation}
\mathbb{E}_G \big(E_g(\ket{G})\big) \leq n-  \mathbb{E}_G \left[\max_{S\subseteq [n]}\big(|S| - \Rank{A_G[S]}\big)\right].
\end{equation}
We can further upper bound this by only maximizing over sets $S$ that are pairwise far away in edit distance\footnote{This is a trick to make the rest of the proof easier. This is also the reason we obtain an exponent $1/4$ instead of the expected $1/2$. We think the proof can be done without this trick (to obtain a tighter bound), but at the cost of a substantial increase in combinatorial complexity.}. Define the set of sets $\mathcal Q$ by dividing $[n]$ up into intervals of size $\sqrt{n}$ (which is an integer by assumption) and taking all sets $S\subset[n]$ that contain $\sqrt{n}/2$ such intervals. This implies that all $S \in \mc Q $ have size $|S| = n/2$, that $|\mathcal Q| = \binom{\sqrt{n}}{\sqrt{n}/2}$, and that all sets in $\mathcal Q$ are pairwise distant (at least $\sqrt{n}$ ) in edit distance. \\

Since $|S| = n/2$ for $S \in \mathcal Q$, we can lower bound the expected maximal rank deficiency (over $\mathcal Q$) by defining the events $E_S(t) = \{\mathrm{Rk}(A)\leq n/2-t\}$; recall \Cref{lem:event_prob}. From the Markov and Bonferroni inequalities (Fact \ref{fact:bonferroni}) we see 
\begin{align}
\mathbb{E}_G \left[\max_{S\in \mathcal Q} \big(|S| - \Rank{A_G[S]} \big) \right] &\geq t \mathbb{P}\bigg(\bigcup_{S\in \mathcal Q} E_S(t)\bigg)\\
&\geq t\,\bigg[ \underbrace{\sum_{S\in \mathcal Q}  \mathbb{P}\big( E_S(t)\big)}_{(1)} - \frac{1}{2}\underbrace{\sum_{S, S'\in \mc Q,\;S\neq S'}\mathbb{P}\big( E_S(t)\cup E_{S'}(t)\big)}_{(2)}\bigg].
\end{align}
We can give a lower bound of the first term (1) using \Cref{lem:event_prob}:
\begin{equation}
(1) = \sum_{S\in \mathcal Q}  \mathbb{P}\big( E_S(n/2 -t)\big)  \geq \frac{1}{4}e^{-2}\binom{\sqrt{n}}{\sqrt{n}/2} e^{-\frac{t^2 +t}{2}}.
\end{equation}

It remains to upper bound the second term (2). To this end, we first recall \Cref{lem:independence intersection} which implies that the joint probability $\mathbb{P}\big( E_S(t)\cup E_{S'}(t)\big)$ only depends on the size of the intersection $I = S\cap S'$. 
The size of this intersection can take values $|I| = 0, \sqrt{n},\ldots, n/2 -\sqrt{n}$. 
Given a set $S \in \mc Q$, for each value of $w \in [0, \sqrt{n}/2 -1]$ there are $\binom {\sqrt n}{w}\binom{\sqrt n}{\sqrt n - w}$ sets $S' \in \mc Q$ satisfying $|S \cap S'| = w \sqrt n$.
Let us choose representatives $S_0 \coloneqq [n/2]$, $S'_w \coloneqq [n/2-\sqrt{n}w, n-\sqrt{n}w]$ with $I_w \coloneqq S_0\cap S'_w$ of size $|I_w| = w\sqrt n $. 
 We can then write
\begin{align}
(2) &= \sum_{S, S'\in \mc Q,\;S\neq S'}\mathbb{P}\big( E_S(t)\cup E_{S'}(t)\big)\\
&= \sum_{S, S'\in \mc Q,\;S\neq S'} \sum_{i,i'=0}^{t}
\mb P\left( \Rank{A[S]} = i \vee \Rank{A[S']} = i'\right)\\
&= \sum_{S, S'\in \mc Q,\;S\neq S'} \sum_{j=0}^{|S \cap S'|} \sum_{i,i'=j}^{t}
\mb P\left( \Rank{A[S]} = i \, | \, \Rank{A[I]} = j\right)\mb P\left( \Rank{A[S]} = i \, | \, \Rank{A[I]} = j\right)\notag\\
&\hspace{10cm} \times\mb P\left( \Rank{A[I]} = j \right)
\end{align}
\begin{align}
&= \binom{\sqrt n }{\sqrt n/2} \sum_{w=0}^{\sqrt n/2 -1} \binom{\sqrt n}{w} \binom{\sqrt n}{\sqrt n/2 - w} \sum_{j=0}^{w\sqrt n/2} \sum_{i,i'=j}^{t}
\mb P\left( \Rank{A[S_0]} = i \, | \, \Rank{A[I_w]} = j\right) \notag\\
&\hspace{4cm} \times \mb P\left( \Rank{A[S_w']} = i' \, | \, \Rank{A[I_w]} = j\right)\mb P\left( \Rank{A[I_w]} = j \right)\\
&= \binom{\sqrt{n}}{\sqrt{n}/2} \sum_{w= 0}^{\sqrt{n}/2 -1} \binom{\sqrt{n}/2}{w}\binom{\sqrt{n}/2}{\sqrt{n}/2 - w} \notag \\
&\hspace{5em}\times \sum_{j=0}^{w\sqrt{n}/2}\bigg(\sum_{i = j}^{n/2-t}\mathbb{P}\big(\Rank{A[S_0]} = i\; \big|\; \Rank{A[I_w]} = j \big)\bigg)^2\mathbb{P}\big(\Rank{A[I_w]} = j \big).
\end{align}
In the last line we used that since $|S_0| = |S'_w|$ we have 
\begin{equation}
\mathbb{P}\big(\Rank{A[S_0]} = i\; \big|\; \Rank{A[I_w]} = j \big) = \mathbb{P}\big(\Rank{A[S'_w]} = i \;\big|\; \Rank{A[I_w]} = j \big).
\end{equation}
Next we use \Cref{fact:rand_adj_rank}, which tells us the rank of $A[I_w]$ is likely nearly maximal. To make this precise we introduce a constant $\alpha\geq 1$ which will be precisely determined later, and split the sum over $j$ into $j< |I_w| - \alpha t$ and $j\geq  |I_w| - \alpha t$. Focusing on this sum we see 
\begin{align}
&\sum_{j=0}^{\lfloor\sqrt{n}w/2\rfloor}\bigg[\sum_{i = j}^{n/2-t}\mathbb{P}\big(\Rank{A[S_0]} = i\; \big|\; \Rank{A[I_w]} = j \big)\bigg]^2\mathbb{P}\big(\Rank{A[I_w]} = j \big)\\
&\leq \sum_{j=\alpha t}^{\lfloor\sqrt{n}w/2\rfloor}\bigg[\sum_{i = j}^{n/2-t}\mathbb{P}\big(\Rank{A[S_0]} = i\; \big|\; \Rank{A[I_w]} = j \big)\bigg]^2 \\
&\hspace{2em}+ e^{2/3}\sum_{j=0}^{\lfloor\sqrt{n}w/2\rfloor-\alpha t- 1}  2^{-\frac{(\lfloor\sqrt{n}w/2\rfloor-j)^2}{2}} 2^{\frac{(\lfloor\sqrt{n}w/2\rfloor-j)}{2}}\bigg[\sum_{i = j}^{n/2-t}\mathbb{P}\big(\Rank{A[S_0]} = i\; \big|\; \Rank{A[I_w]} = j \big)\bigg]^2\notag\\
&\leq \sum_{j=\alpha t}^{\lfloor\sqrt{n}w/2\rfloor}\bigg[\sum_{i = j}^{n/2-t}\mathbb{P}\big(\Rank{A[S_0]} = i\; \big|\; \Rank{A[I_w]} = j \big)\bigg]^2\notag\\
& \hspace{5cm}+ e^{2/3}\sum_{j=0}^{\lfloor\sqrt{n}w/2\rfloor-\alpha t- 1}  2^{-\frac{(\lfloor\sqrt{n}w/2\rfloor-j)^2}{2}} 2^{\frac{(\lfloor\sqrt{n}w/2\rfloor-j)}{2}}\\
&\leq \sum_{j=\alpha t}^{\lfloor\sqrt{n}w/2\rfloor}\bigg[\sum_{i = j}^{n/2-t}\mathbb{P}\big(\Rank{A[S_0]} = i\; \big|\; \Rank{A[I_w]} = j \big)\bigg]^2\notag\\
& \hspace{5cm}+ e^{2/3}\sum_{j=0}^{\lfloor\sqrt{n}w/2\rfloor-\alpha t- 1}  2^{-\frac{(\lfloor\sqrt{n}w/2\rfloor-j)^2}{2}} 2^{\frac{(\lfloor\sqrt{n}w/2\rfloor-j)}{2}},
\end{align}
where we've used \Cref{fact:rand_adj_rank} 
to upper bound $ \mathbb{P}\big(\Rank{A[I]} = j \big)$ in the second term, and have trivially bounded $\mathbb{P}\big(\Rank{A[I]} = j \big)\leq 1$ in the first. 
It remains to use \Cref{lem:cond_upper_bound} to bound the first term. We obtain
\begin{multline}
(2) \leq \binom{\sqrt{n}}{\sqrt{n}/2} \sum_{w= 0}^{\sqrt{n}/2 -1} \binom{\sqrt{n}/2}{w}\binom{\sqrt{n}/2}{\sqrt{n}/2 -w}\\ \times \bigg[ \sum_{j=\alpha t}^{w \sqrt n /2}\Big(\sum_{i = j}^{n/2-t} 2\cdot 2^{-\frac{(2i)^2}{2}+ \frac{(2i)}{2}} +7\cdot2^{2i\alpha} \rho^{(\sqrt{n}/2 - w)\sqrt{n} }\Big)^2 + 2\cdot2^{-\frac{(\alpha t -1)^2}{2}} \bigg].
\end{multline}
Setting $t=5\sqrt{\log\Big[\binom{\sqrt{n}}{\sqrt{n}/2}\Big]}$ and $\alpha = 10$ we can evaluate the resulting expression (packing all numbers into a constant $C>0$) and obtain
\begin{multline}
\mathbb{E}_G \big(\max_{S\in \mathcal Q} \frac{n}{2} - \Rank{A_G[S]} \big)\ge  \\\binom{\sqrt{n}}{\sqrt{n}/2} 
\frac{1}{4}e^{-2} e^{-\frac{t^2 +t}{2}} - 
 \binom{\sqrt{n}}{\sqrt{n}/2} \sum_{w= 0}^{\sqrt{n}/2 -1} \binom{\sqrt{n}/2}{w}\binom{\sqrt{n}/2}{\sqrt{n}/2 -w} \Bigg[\sum_{j=\alpha t}^{\lfloor\sqrt{n}w/2\rfloor} \Big(\sum_{i = j}^{n/2-t} 2\cdot 2^{-\frac{(2i)^2}{2}+ \frac{(2i)}{2}}\\
  +7\cdot2^{2i\alpha} \rho^{(\sqrt{n}/2 - w)\sqrt{n} }\Big)^2 + 2\cdot2^{-\frac{(\alpha t -1)^2}{2}} \Bigg] \\
\geq C \sqrt{\log\bigg[\binom{\sqrt{n}}{\sqrt{n}/2}\bigg]} = \Omega \big(n^{(1/4)}/\log(n)\big),
\end{multline}
which is what we set out to prove.
\end{proof}
\subsection{Structured simulation algorithms}\label{subsec:structured}
The upper bound on the geometric entanglement derived above is strong enough to break the simulation algorithm derived in \cite{gross_most_2009}. However, better simulation algorithms might be found by exploiting the extra structure that graph states provide. In particular, we can think of any MBQC procedure as a sequence of measurements in the eigenbases of the $X,Y, X-Y,X+Y$ operators \cite{van2006universal}. On a graph state the first three measurements can be simulated classically in polynomial time. We can thus envision an improved simulation algorithm where we classically simulate the $X,Y,Z$ measurements and simulate the $X+Y,X-Y$ measurements by coin flips. The efficacy of this algorithm depends critically on how the overlap with  a random graph state fluctuates with respect to tensor products of the eigenstates of $X+Y,X-Y$. As these states are highly magical, it is possible that their overlap with stabilizer states fluctuates much less than the maximal overlap with arbitrary stabilizer states (which as we saw above, is dominated by contributions from product \emph{stabilizer} states). We believe that this is not the case (and thus that this algorithm does not work), but can not prove it as of yet. We think this is an interesting question in its own right, so we leave it as a conjecture. 
\begin{conjecture}
Labeling the set of eigenstates of $X+Y,X-Y$ as $\mathcal{T}$, there exist  $c,C>0$ s.t.
\begin{equation}
\mathbb{P}\left[ \max_{\ket{\alpha}\in \mc{T}\tn{n}} |\langle \alpha \ket{G}|^2 \geq 2^{-n+ n^{c}} \right]\geq C,
\end{equation}
with the probability taken uniformly over graph states.
\end{conjecture}

\bibliographystyle{alpha}
\bibliography{doms_library,graphstates}

\end{document}